\documentclass[12pt,american]{article}

\usepackage{amstext}
\usepackage{array}
\usepackage{babel}
\usepackage{cancel}
\usepackage{comment}
\usepackage{float}
\usepackage[T1]{fontenc}
\usepackage[a4paper]{geometry}
\usepackage{graphicx}
\usepackage[utf8]{inputenc}
\usepackage{lmodern}

\usepackage{mathtools}
\usepackage{mathdots}
\usepackage[numbers,sort]{natbib}
\usepackage{rotating}
\usepackage{setspace}
\usepackage{stackrel}
\usepackage{subcaption}
\usepackage{stmaryrd}
\usepackage{subscript}
\usepackage{textcomp}
\usepackage{units}
\usepackage[hyphens]{url}
\usepackage{verbatim}
\usepackage{xcolor}

\usepackage[unicode=true,pdfusetitle,
 bookmarks=true,bookmarksnumbered=false,bookmarksopen=false,
 breaklinks=false,pdfborder={0 0 0},backref=false,colorlinks=false]
 {hyperref}
\usepackage{bibunits}

\makeatletter

\usepackage{eurosym}

\pdfpageheight\paperheight
\pdfpagewidth\paperwidth

\providecommand{\tabularnewline}{\\}

\makeatother

\begin{document}

\begin{bibunit}[IEEEtranSN]

\title{Optimal supply chains and power sector benefits of green hydrogen}

\author{
    Fabian Stöckl \\
    \small German Institute for Economic Research (DIW Berlin), Germany, and \\[-5pt] 
    \small Technische Universität Berlin, Germany
    \and
    Wolf-Peter Schill \\
    \small German Institute for Economic Research (DIW Berlin), Germany, and \\[-5pt]
    \small climate \& Energy College, Energy Transition Hub, University of Melbourne, Australia \\[-5pt]
    \small Corresponding author, \texttt{wschill@diw.de}
    \and
    Alexander Zerrahn \\
    \small German Institute for Economic Research (DIW Berlin), Germany
}

\maketitle
\begin{center}
\vspace{-0.6cm}

\par\end{center}
\renewcommand{\abstractname}{Summary}
\begin{abstract}
    \noindent Green hydrogen can help to decarbonize parts of the transportation sector, but its power sector interactions are not well understood so far. It may contribute to integrating variable renewable energy sources if production is sufficiently flexible in time. Using an open-source co-optimization model of the power sector and four options for supplying hydrogen at German filling stations, we find a trade-off between energy efficiency and temporal flexibility. For lower shares of renewables and hydrogen, more energy-efficient and less flexible small-scale on-site electrolysis is optimal. For higher shares of renewables and/or hydrogen, more flexible but less energy-efficient large-scale hydrogen supply chains gain importance, as they allow to temporally disentangle hydrogen production from demand via storage. Liquid hydrogen emerges as particularly beneficial, followed by liquid organic hydrogen carriers and gaseous hydrogen. Large-scale hydrogen supply chains can deliver substantial power sector benefits, mainly through reduced renewable curtailment. Energy modelers and system planners should consider the distinct flexibility characteristics of hydrogen supply chains in more detail when assessing the role of green hydrogen in future energy transition scenarios. We also propose two alternative cost and emission metrics which could be useful in future analyses.
\end{abstract}


\thispagestyle{empty}

\newpage{}


\setcounter{page}{1}

\section{Introduction\label{sec: Introduction-P2H2}}

The increasing use of renewable energy sources in all end-use sectors is a main strategy to reduce greenhouse gas emissions~\citep{DeConinck2018}. This not only applies to the power sector, but also to other sectors such as transportation. There, energy demand may be satisfied either directly by renewable electricity or indirectly by hydrogen and derived synthetic fuels produced with renewable electricity~\citep{Armaroli2011,Yan2019, Staffell2019, Brynolf2018, DeLuna2019}. The potential role of hydrogen-based electrification for deep decarbonization is widely acknowledged~\citep{Jacobson2017, Luderer2018a,Hanley2018,NatureEnergy2016a}. 

Yet, a central aspect is less understood so far: how hydrogen-based electrification interacts with the power sector. Hydrogen supply chains use different types of storage, which allow to temporally disentangle electricity demand for hydrogen production from the time profile of final hydrogen demand. Similar to other flexibility options in the power sector, such as load shifting or electricity storage, this increases the \textit{temporal flexibility} of the power sector. Such flexibility can help make better use of variable renewable energy from wind and solar PV \citep{Kondziella2016a,Bird2013a}. This, in turn, impacts the optimal electricity generation and storage capacities in the power sector, their hourly use, carbon emissions, and costs. Yet more flexible hydrogen supply chains may be less energy-efficient as they incur more conversion steps \citep{Reuss2017av,Yang2007a}. Thus, the overall power system impacts of different hydrogen supply chains, considering both their flexibility and energy efficiency characteristics, are a priori unclear.

We address this research gap on the power sector interactions of green hydrogen by investigating different supply chains of hydrogen for road-based passenger mobility for future scenarios with high shares of variable renewable electricity. Specifically, we determine least-cost options for the supply of electrolysis-based hydrogen at filling stations, while explicitly considering how they interact with the power sector. To this end, we use an open-source cost-minimization model with a technology-rich well-to-tank perspective that co-optimizes the power sector and four relevant hydrogen supply chains derived from the literature: small-scale on-site electrolysis at the filling station as well as three large-scale hydrogen production and distribution options. 

As outlined in more detail in Section~\ref{sec: lit rev}, many previous power sector analyses that include hydrogen for mobility lack detail with respect to the representation of hydrogen production and distribution options~\citep{Breyer2019a,Bogdanov2021a,Gils2017a,Brown2018a,Oldenbroek2021a}. In contrast, studies that include more techno-economic details of supply chains for hydrogen mobility often rely on exogenous electricity price inputs, include only rudimentary power sectors, tie hydrogen production to the availability of surplus electricity generation, and/or are restricted to a single supply chain~\citep{Emonts2019a,Glenk2019,Reuss2017av,Runge2019av,Robinius2017a,Welder2018a,Yang2007a,Samsatli2016a}. Yet, none of these studies examines the interactions between hydrogen supply chains and power sectors with high shares of renewable energy sources in detail.

In this paper, we develop and apply an integrated hydrogen and power sector model to fill this gap in the literature. It minimizes overall system costs by endogenously optimizing electricity generation and storage capacities, their hourly dispatch, as well as capacity and hourly use of hydrogen supply chains. We parametrize our model to a 2030 setting for Germany. The insights derived for this case study should also be of interest to a range of other countries (see Section~\ref{sec: Discussion-P2H2}). Germany's power sector is the largest in Europe. Traditionally, it has been dominated by thermal power plants, and is now increasingly shifting towards variable renewable energy sources, as dispatchable renewable sources such as hydro or geothermal energy are limited. Since the first version of the Renewable Energy Sources Act (EEG) entered into force in 2000, the German government repeatedly committed itself to an ambitious expansion of renewable energy sources \citep{Hake2015a}. This has put Germany among the global front-runner countries in terms of variable renewable energy use \citep{REN21a}. Recently, the German government also substantially increased its ambition to use green hydrogen and aims to become a major supplier of green hydrogen technologies \citep{BMWi2020}.


\section{Literature review}\label{sec: lit rev}

The existing literature covering the use of hydrogen in the mobility sector can be roughly divided into two groups based on their focus. The first group of analyses focuses on decarbonizing the power sector or even the whole economy, but models the use of hydrogen only in a very stylized way. The second group focuses on a detailed techno-economic representation of different hydrogen production and distribution schemes, but not on their interaction with the power sector.

The first group, which focuses on decarbonization, comprises, for instance, an analysis of \citeauthor{Breyer2019a} \citep{Breyer2019a}. The authors model a generic H\textsubscript{2} demand for mobility (road, marine, aviation, rail) within a worldwide \unit[100]{\%} renewable energy setting for 2050. They find a share of hydrogen accounting for about \unit[25]{\%} of total energy demand in the transportation sector. Yet, as the study covers energy systems on a global scale, it does not provide great temporal and technological detail of hydrogen production and distribution.
The same is true for \citeauthor{Bogdanov2021a} \citep{Bogdanov2021a}. There, the authors apply a similar model to the case of Kazakhstan, an extreme example with unfavorable climatic conditions and an energy-intensive industry that make decarbonization with renewable energy sources more challenging than elsewhere.
Similarly, \citeauthor{Gils2017a} \citep{Gils2017a} develop a \unit[100]{\%} renewable energy scenario for the Canary Islands, with hydrogen powering between \unit[37]{\%} and \unit[75]{\%} of road transportation. The authors find that the use of hydrogen may provide additional flexibility and facilitate the integration of high shares of renewables. However, their model only features a stylized hydrogen sector with a generic storage option. 
\citeauthor{Brown2018a} \citep{Brown2018a} compare the effect of increased coupling of the power, heat, and transportation sectors vis-à-vis an extension of electricity transmission networks for high shares of renewables in Europe. Modeling different fleets of fuel-cell electric vehicles (FCEV), the authors find that the availability of large-scale storage can make hydrogen an important flexibility option in the power sector. Yet, due to its high temporal and spatial resolution, only a simplified hydrogen supply system is modeled.
Finally, \citeauthor{Oldenbroek2021a} \citep{Oldenbroek2021a} analyze \unit[100]{\%} renewable energy scenarios with fuel-cell electric vehicles that also feature vehicle-to-grid support for several European countries. They find that the backup power provided by a share of \unit[50]{\%} of cars being grid-connected FCEVs is sufficient to balance the power system. However, the hydrogen sector modeling does not account for different production and distribution chains.
While all these studies have their merits, none provides sufficient techno-economic hydrogen sector details for an in-depth analysis of how different hydrogen production and distribution options interact with the power sector.

In contrast to the studies mentioned above, the second group of literature focuses on highly detailed representations of the hydrogen sector. For instance, \citeauthor{Welder2018a} \citep{Welder2018a} and \citeauthor{Samsatli2016a} \citep{Samsatli2016a} analyze hydrogen-to-mobility for Germany and Great Britain, respectively, but do not connect hydrogen production to the power system. Instead, electricity demand for electrolysis is covered by a wind power capacity built and used exclusively for that purpose. Moreover, the transmission of hydrogen is restricted to pipelines. The authors find H\textsubscript{2} to be cost-competitive with fossil fuels in the transportation sector \citep{Welder2018a}, and that all of Great Britain's domestic transport can be supplied by onshore wind-powered hydrogen production \citep{Samsatli2016a}. 
In a similar study, \citeauthor{Robinius2017a} \citep{Robinius2017a} restrict electrolysis to be powered only by renewable surplus energy, which is derived for a predetermined capacity in a German 2050 scenario. Again, hydrogen is assumed to be transported to filling stations via pipelines. The authors find that renewable surplus electricity would be sufficient to serve Germany's hydrogen demand for mobility. However, the research design neglects the effects hydrogen production may have on the power sector.
Studying hydrogen-for-mobility pathways in Germany, \citeauthor{Emonts2019a} \citep{Emonts2019a} consider various options for hydrogen distribution and storage. Again, the electricity demand for water-electrolysis is constrained to be served by renewable surplus generation within a predetermined future energy system. The authors identify pipeline transmission as cost-optimal for large demands of hydrogen, while transportation via trucks is more favorable for lower demands.
Taking the view of a wind turbine operator, \citeauthor{Glenk2019} \citep{Glenk2019} compare the grid feed-in of wind energy with its alternative use for hydrogen production. They find that hydrogen production is currently more profitable if it can be sold at prices of ~\unit[3.23]{€/kg} (Germany) and ~\unit[3.53]{\$/kg} (Texas). The analysis abstracts from power sector modeling and instead draws on past electricity prices.
\citeauthor{Yang2007a} \citep{Yang2007a} and \citeauthor{Reuss2017av} \citep{Reuss2017av} compare several highly detailed hydrogen production and distribution chains. They find that the cost-optimal supply chain mainly depends on the average transportation distance and on overall hydrogen demand. Yet again, their analyses rely on exogenous electricity price assumptions.
In contrast, \citeauthor{Runge2019av} \citep{Samsatli2016a} use an electricity market model to derive hourly electricity prices for different market designs to analyze the costs of hydrogen supply based on liquid organic hydrogen carriers at German filling stations. Still, their model setup covers only the effect of electricity prices on the hydrogen sector, but not the feedback in the other direction. While all analyses of the second group represent the hydrogen sector with high techno-economic detail, they do not allow for investigating its interaction with the power sector.

There are only a few studies that explicitly account for such interactions between the two sectors. These are located between the two groups sketched above and conceptually closest to our work. Despite uncovering important new insights, in general, they are characterized by an incomplete co-optimization of sectors or a lack of detail in the representation of hydrogen production and distribution.For instance, \citeauthor{Michalski2017a} \citep{Michalski2017a} investigate the impact of different availabilities of hydrogen mobility infrastructure on German power plant dispatch. In contrast to our study, hydrogen infrastructure, as well as large parts of electricity generation capacities, are exogenously set and not the result of a co-optimization of the two sectors. Moreover, while the authors allow for both centrally produced hydrogen close to large underground storage (salt caverns) and on-site production at filling stations, the share of either type is not endogenous, but fixed. Results indicate that large-scale cavern storage allows for flexible deployment of electrolysis capacities, thus reducing the curtailment of wind and solar power.
In another study, \citeauthor{Rose2020a} \citep{Rose2020a} optimize the spatial distribution and storage size of hydrogen filling stations for heavy-duty trucks in Germany. They show that co-optimization with respect to total system cost, i.e.,~including the power sector and local grid restrictions, can reduce costs of hydrogen provision by up to \unit[10]{\%} compared to a non-optimized spatial distribution pattern. This cost reduction is mainly driven by lower electricity costs, as hydrogen production is better aligned with the sector's flexibility needs. However, their analysis only considers on-site electrolysis at filling stations and thus cannot account for potential trade-offs between different hydrogen production and distribution options.
Finally, \citeauthor{Zhang2020a} \citep{Zhang2020a} investigate the effect of flexible hydrogen production for mobility on the Western U.S.~power system. They find a trade-off between the benefits of flexible hydrogen production for the power system operation and the expenditures for respective infrastructure costs, i.e.,~additional electrolyzer and storage capacity. In their model, a minimum of total costs is reached for slightly oversized electrolyzers in the range of 11 to \unit[25]{\%} compared to a scenario with flat production and a capacity factor of \unit[100]{\%}. However, their analysis is based on a very stylized hydrogen sector and lacks a representation of the specific characteristics of different hydrogen distribution pathways. Moreover, the authors focus on the dispatch of existing generation capacities and neglect the effects of hydrogen on optimal capacity expansion.

Overall, the review of the existing literature reveals a lack of detail in the representation of either the hydrogen or the power sector, and/or a missing co-optimization of both sectors. This impedes a thorough analysis of potential benefits and challenges related to the interaction of variable renewables and different hydrogen production and distribution options. Our analysis adds to the literature by providing, to the best of our knowledge, the first full co-optimization of different hydrogen supply chains and the power sector. Thereby, we consider a range of scenarios in which we systematically vary assumptions on future hydrogen demand and the share of renewable energy sources. Specifically, we investigate the trade-off between energy efficiency and temporal flexibility for different hydrogen supply chains, and how it interacts with optimal capacity and dispatch outcomes in the power sector.


\section{Model and scenarios}\label{sec: Model}

\subsection{The open-source power sector model DIETER}
We use the established open-source power sector model DIETER. Different versions of this model have been previously used for analyzing aspects of renewable energy integration with a focus on utility-scale energy storage ~\citep{Zerrahn2017,Schill2018,Schill2020a}, decentralized storage related to prosumers\cite{Say2020a,Guenther2021}, and power-to-heat options \cite{Schill2020}. The model includes several features essential for meaningful analyses of integrating variable renewable energy sources, in particular a sufficient temporal resolution \citep{Ringkob2018a,Pfenninger2014a} and a detailed modeling of energy storage \citep{Bistline2020a}. For transparency and reproducibility~\citep{Pfenninger2017}, the source code, input data, and a complete documentation of the model version used here are available under a permissive open-source license in a public repository~\citep{Stoeckl2020v} (see also \url{www.diw.de/dieter}).

The model minimizes the total system costs of providing electricity and hydrogen. The objective function comprises annualized investment costs and hourly variable costs of electricity generation and storage technologies, electrolysis, as well as storage, conversion, and transportation of hydrogen. The main model inputs are availability and cost parameters for all technologies as well as hourly time series of electricity demand, hydrogen demand, and renewable capacity factors. The main decision variables are capacities in the power and hydrogen sectors as well as their hourly use. The optimization is subject to constraints, including market balances for electricity and hydrogen that equate supply and demand in each hour, capacity limits for generation and investment, and a minimum share of renewable energy in electricity supply. The model determines a long-run first-best equilibrium benchmark for a frictionless market. Assuming perfect foresight, DIETER is solved for all consecutive hours of an entire year, thereby capturing the variability of renewable energy sources. Model outputs comprise system costs, optimal capacities and their hourly use, and derived metrics such as emission intensities.

The electricity demands of various processes along the hydrogen supply chains enter the model's energy balance. This includes electricity used for hydrogen production, processing, and distribution facilities. Depending on the conversion steps along the supply chain, the four options differ in how much electricity is required overall, and at which stage of the process (for an illustration, see Section~\ref{sub: hydrogen sector data}). All costs for hydrogen-related investments enter the model's objective function.

As we aim to derive general insights on temporal flexibility, we abstract from an explicit representation of idiosyncratic spatial aspects and electricity network constraints. Moreover, to keep the analysis tractable, the DIETER version used here has no explicit representation of electricity transmission, focuses on Germany only, and abstracts from balancing within the European interconnection. We also do not use some features of the original model, such as demand-side flexibility beyond the hydrogen sector.


\subsection{Four hydrogen supply chains}
The hydrogen sector is modeled with a well-to-tank perspective. It includes four options to provide filling stations with hydrogen: small-scale on-site electrolysis directly at the filling station, and three more centralized large-scale options, where H\textsubscript{2} is delivered by trailers (Figure~\ref{fig: Schaubild}). The three centralized supply chains are adapted from previous analyses of hydrogen production and distribution \citep{Reuss2017av,Yang2007a} and are characterized by the availability of large-scale hydrogen generation and storage. They mainly differ with respect to the form in which hydrogen is stored: gaseous hydrogen (GH$_2$), liquid hydrogen (LH$_2$), or bound to a liquid organic hydrogen carrier (LOHC, see~\cite{Preuster2017v}). In contrast, on-site hydrogen production, which leans on \citep{Michalski2017a}, comes with only limited amounts of hydrogen storage in high-pressure gas tanks, which is motivated by space and security reasons. For our analysis, only one supply chain can be selected per filling station.

Small-scale on-site hydrogen production is restricted to proton exchange membrane (PEM) water electrolysis, which is superior to alkaline (ALK) electrolysis in several dimensions relevant for small-scale on-site production, including higher load flexibility~\citep{Mittelsteadt2015a}, a lower footprint~\citep{Mittelsteadt2015a}, and easier handling~\citep{Linde2016a}. Locally produced hydrogen is immediately compressed and stored at~$700$-\unit[950]{bar} in high-pressure vessels at the filling station. The same high-pressure storage and dispensing installations are also present in the large-scale supply chains.

For large-scale hydrogen production, we consider both ALK and PEM electrolysis. As the large-scale options allow for bulk hydrogen storage, they provide greater temporal flexibility compared to the small-scale on-site option, which only comes with a short-term buffer storage at the filling station. Hydrogen from electrolysis is either compressed and stored at the production site at up to~\unit[250]{bar} (GH$_2$), liquefied and stored in insulated tanks (LH$_2$), or bound to a liquid organic hydrogen carrier (LOHC) in an exothermic hydrogenation reaction and stored in simple tanks. As LOHC, we assume dibenzyltoluene; see~\cite{Eypasch2017av} for an exposition. GH$_2$ and LOHC can be stored without losses; LH$_2$ suffers from a boil-off of~$\sim$\,\unit[0.2]{\%} per day ($\sim$\,\unit[52]{\%} per year), which lowers its potential for long-term H$_2$ storage. For GH$_2$, hydrogen may also be directly prepared for transportation after production, bypassing production-site storage. Investments in storage capacity at large-scale production sites are unrestricted. Due to minimum filling level requirements, usable storage capacities can be lower than nominal capacities.

For transportation, hydrogen is taken from the respective storage at the large-scale production site, re-compressed (if necessary), and transported (time-consuming) in special trailers to the filling stations.


At filling stations, GH$_2$ from large-scale electrolysis is either re-compressed and stored at up to~\unit[250]{bar} or directly compressed to~\unit[950]{bar} for the high-pressure buffer storage (bypass option). LH$_2$ and LOHC are first stored in unconverted form, where boil-off for LH$_{2}$ is slightly higher at the filling station than at the large-scale production site ($\sim$\,\unit[0.4]{\%} per day or~$\sim$\,\unit[77]{\%} per year). Spatial limitations and security aspects restrict these storage capacities to two trailer-loads for all three large-scale supply chains. LH$_2$ is then cryo-compressed and evaporated, and LOHC dehydrogenated and compressed to be stored in gaseous form at up to~\unit[950]{bar} in high-pressure vessels used as a buffer for dispensing. High-pressure storage is limited to~\unit[300]{kg} (one~\unit[20]{ft} container with tubes~\citep{HexagonComposites2016a}).

\begin{center}
\begin{figure}[H]
\begin{centering}
\includegraphics[width=0.95\textwidth]{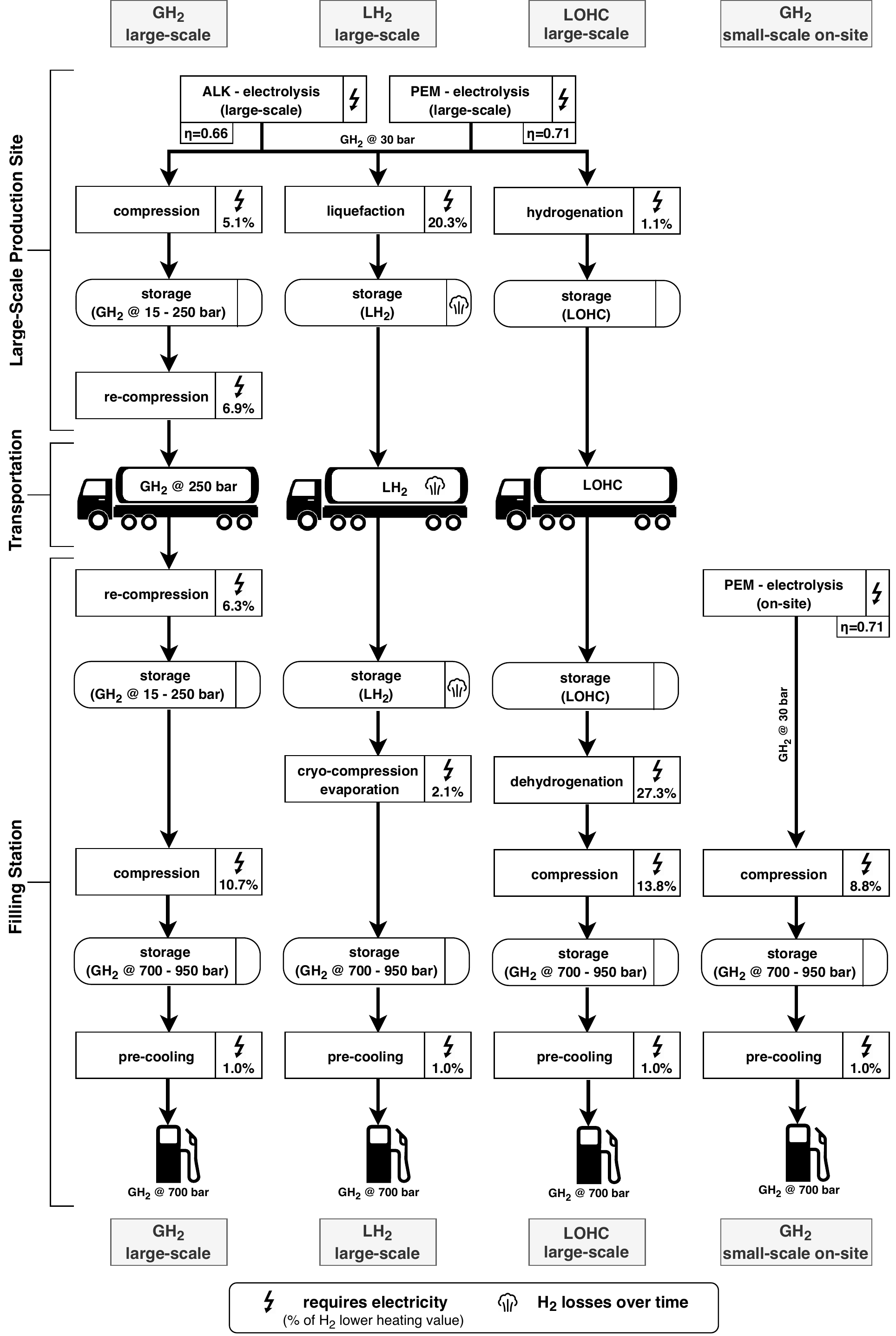} 
\par\end{centering}
\protect\caption{\label{fig: Schaubild} Large-scale and small-scale on-site supply chains with specific production, processing, transportation, and storage requirements.}
\end{figure}
\par\end{center}



\subsection{Renewable energy share and hydrogen demand scenarios}
Twelve scenarios vary the share of renewable energy sources in electricity generation between~$65$-\unit[80]{\%} in five percentage point increments, and the demand for hydrogen between~$0$, $5$, $10$, and~\unit[25]{\%} of private and public road-based passenger vehicle energy demand. A renewable share of~\unit[65]{\%} exactly matches the current German government's target for~2030. Larger shares reflect higher ambition levels, which Germany aims for beyond 2030, and are required for achieving more progressive climate targets. We abstract from modeling renewable shares beyond~\unit[80]{\%} here, as these appear to be more plausible in longer-term settings in which other sector coupling technologies and the reconversion of hydrogen to electricity would also become more relevant. Note that increasing the share of renewable energy beyond~\unit[65]{\%} requires additional deployment of variable solar PV or wind power capacity, as the potentials of dispatchable hydro and bioenergy sources are already fully realized.

Annual hydrogen demands are~$9.1$, $18.1$, and~\unit[45.3]{TWh$_{H_{2}}$} at the filling stations, representing different potential future market penetrations of hydrogen-electric mobility. This relates to~\unit[5]{\%}, \unit[10]{\%}, or~\unit[25]{\%} of road-based passenger traffic in Germany (compare Section~\ref{sub: hydrogen demand data}). These hydrogen demands substantially exceed those of the fleet of fuel cell electric vehicles that can be reasonably expected in 2030. Yet, the demand levels used here allow for interesting insights in settings where hydrogen demand is non-negligible from a power sector perspective. For clarity, we abstract from the provision of hydrogen for other purposes than mobility.

For each renewable energy share and hydrogen demand scenario, we combine the small-scale on-site hydrogen supply option with either of the three large-scale options. This results in three distinct combinations of options per scenario. Due to path dependencies and technology specialization, we do not expect parallel infrastructures for large-scale technologies to emerge in a plausible future setting.


\section{Results\label{sec: Results}}


\subsection{Optimal hydrogen supply chains depend on renewable penetration and hydrogen demand\label{sub: optimal h2 chains}}

Figure~\ref{fig: 12 Panel Graph} shows the cost-minimal combinations of small-scale on-site (OS) and large-scale hydrogen supply chains for the~$12$~scenarios with hydrogen demand. We denote the resulting renewables-demand scenarios as~$Res65$-$Dem5$, $Res65$-$Dem10$, and so on. The Figure also shows the Additional System Costs of Hydrogen (ASCH, see also Section~\ref{sub: Metrics}), defined as difference in total system costs between a scenario that includes hydrogen and the respective baseline without hydrogen demand, related to total hydrogen supply. 

\begin{center}
\begin{figure}[H]
\begin{centering}
\includegraphics[width=0.80\textwidth]{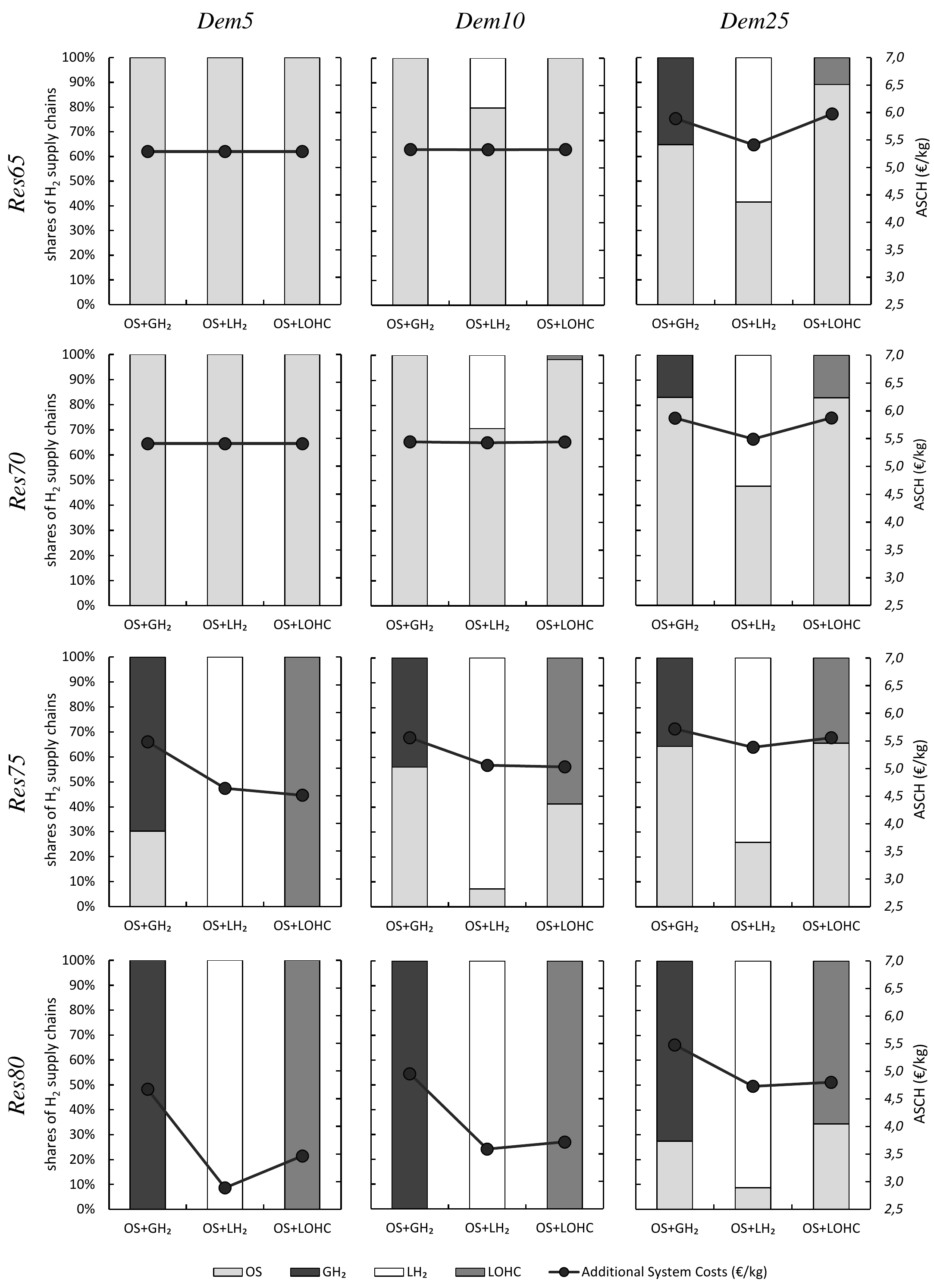} 
\par\end{centering}
\protect\caption{\label{fig: 12 Panel Graph}Optimal combinations of small-scale on-site (OS) and large-scale hydrogen supply chains and Additional System Costs of Hydrogen (ASCH) for the~$12$ scenarios. Starting from the top left panel, the share of renewable energy sources increases to the bottom, and the demand for hydrogen increases to the right.}
\end{figure}
\par\end{center}

For combinations of relatively low shares of renewable energy sources ($65$-\unit[70]{\%}) and hydrogen demand ($5$-\unit[10]{\%} of road-based passenger traffic), small-scale electrolysis is the least-cost option. That is, the energy efficiency benefits of on-site electrolysis prevail over the flexibility benefits of large-scale options. Large-scale supply chains are increasingly part of the optimal solution for higher shares of renewables or greater hydrogen demand. In these scenarios, the flexibility they offer becomes more valuable. Among the three large-scale options, liquid hydrogen tends to have the highest shares in the optimal solution.

Comparing the Additional System Costs of Hydrogen, the solutions that include compressed gaseous hydrogen are always dominated by liquid hydrogen and often also by LOHC. This is because GH$_{2}$, while energy efficient, incurs comparably high storage and transportation costs (see Section~\ref{sub: hydrogen sector data}). In contrast, solutions that include LH$_{2}$ lead to the lowest ASCH in most scenarios with high renewable shares ($75$-\unit[80]{\%}) or high hydrogen demand (\unit[25]{\%}). In general, solutions that include LH$_{2}$ or LOHC often lead to relatively similar cost outcomes. Yet, this is driven by different underlying mechanisms. LH$_{2}$ is overall more energy efficient; LOHC offers higher temporal flexibility due to cheap storage, yet requires substantial amounts of electricity for the dehydrogenation process at the filling station  (see Sections~\ref{sub: utilization patterns} and~\ref{sub: hydrogen sector data}).

Further, the Additional System Costs of Hydrogen generally increase with hydrogen demand and decrease with the share of renewable energy sources, mainly reflecting the availability of cheap renewable surplus energy (see Section~\ref{sub: power sector outcomes}).


\subsection{Use patterns of hydrogen production and storage indicate differences in temporal flexibility\label{sub: utilization patterns}}

Differences in hydrogen storage capabilities as well as the level and timing of electricity demand (Section~\ref{sub: hydrogen sector data}) lead to very different utilization patterns of the four hydrogen supply chains. We illustrate this for the optimal combination of temporally inflexible small-scale electrolysis and more flexible LH$_{2}$ in the $Res80$-$Dem25$~scenario.

\begin{center}
\begin{figure}[H]
\begin{centering}
\subcaptionbox{\label{fig: LH2 in detail - a}Large-scale electrolysis (PEM\,\&\,ALK)}{\includegraphics[width=7cm]{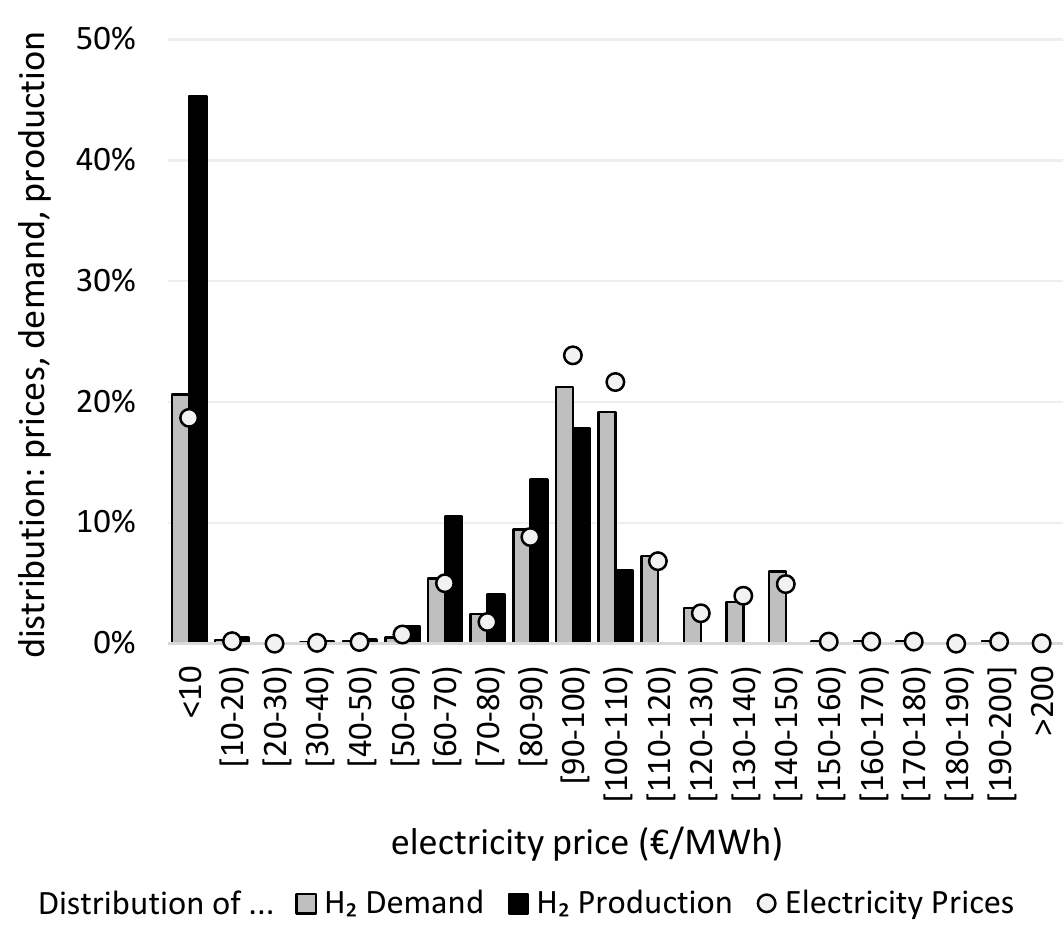}}~~~~~\subcaptionbox{\label{fig: LH2 in detail - b}Small-scale~on-site~electrolysis~(PEM)}{\includegraphics[width=6.9cm]{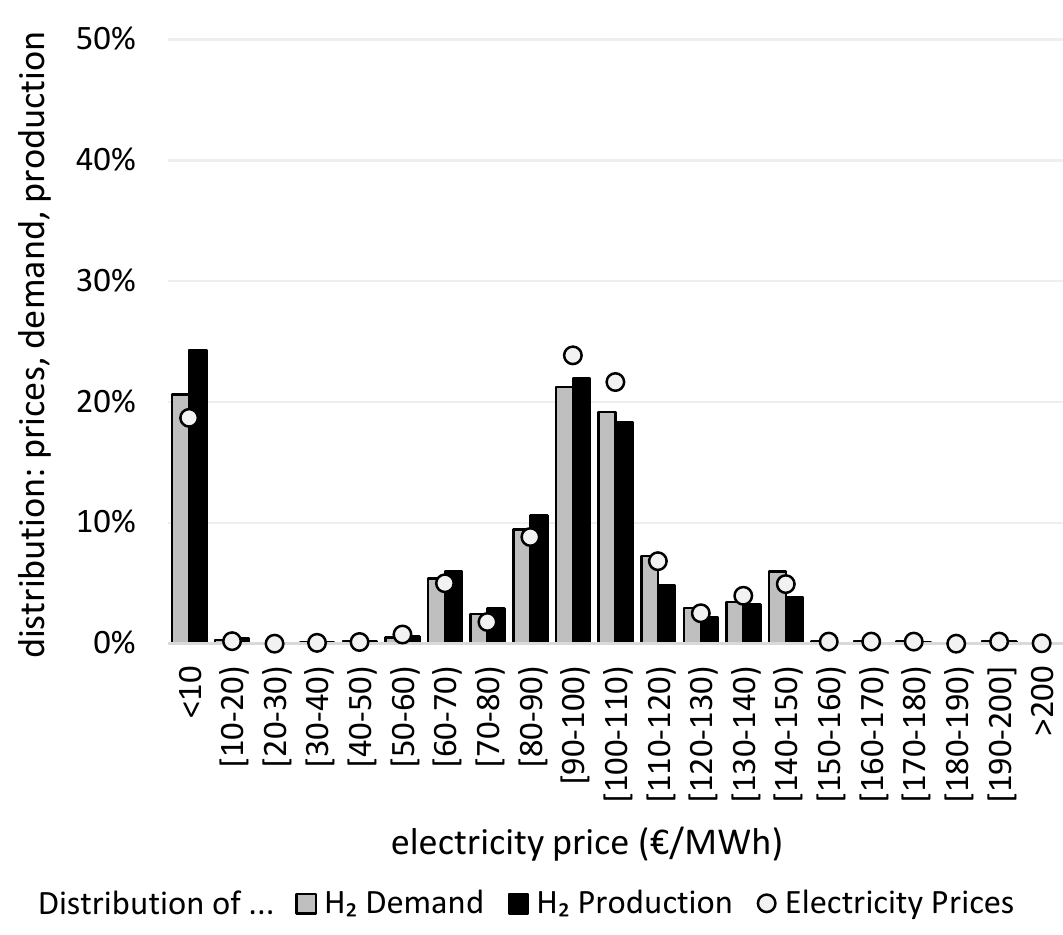}}
\par\end{centering}
\centering{}\caption{\label{fig: LH2 in detail}Distribution of hydrogen production, hydrogen demand, and electricity prices, exemplary for OS+LH$_{2}$ in scenario~$Res80$-$Dem25$}
\end{figure}
\par\end{center}

Figure~\ref{fig: LH2 in detail - a} shows that LH$_{2}$ allows to temporally disentangle hydrogen production from demand. On average, production is high during hours when (renewable) electricity is abundant and, thus, cheap. These are not necessarily hours of high hydrogen demand. At the filling station, dispensing LH$_{2}$ on time requires only little electricity. Vice versa, large-scale hydrogen production is low during hours of high prices. In contrast, on-site electrolysis only includes a small high-pressure buffer storage and needs to produce almost on demand (Figure~\ref{fig: LH2 in detail - b}). Thus, through greater temporal flexibility, LH$_{2}$ allows to exploit phases of high renewable electricity supply and accordingly low electricity prices, which can overcompensate the overall higher electricity demand. Comparable production patterns also emerge for the other two large-scale supply chains GH$_2$ and LOHC.

The capacities of production-site hydrogen storage and its hourly use vary substantially across the three large-scale options (Figure~\ref{fig: Utilization of mass storage}). LOHC has the highest overall storage capacity and a strongly seasonal use pattern. In contrast, GH$_2$ has a much smaller storage capacity and a pronounced short-term storage pattern. LH$_2$ storage is in between. Capacity deployment of GH$_2$ storage is small because of its relatively high specific investment costs. This changes in a sensitivity with cheap cavern storage (see Section~\ref{sub: Sensitivity - Cavern Storage for GH2}). For LH$_2$, storage investment costs are much lower, yet investment costs for liquefaction plants are high, impeding investments in larger LH$_2$ production capacities. LH$_2$ storage is also subject to a small, but relevant boil-off, which makes it less suitable for long-term storage. For LOHC, both investment costs for storage and hydrogenation plants are relatively low and investments, accordingly, high. As there is also no boil-off, LOHC storage is used for seasonal balancing.
\vspace{-0.2cm}
\begin{center}
\begin{figure}[H]
\begin{centering}
\subcaptionbox{\label{fig: Utilization of mass storage - GH2}GH$_{2}$ storage}{\includegraphics[width=4.3cm]{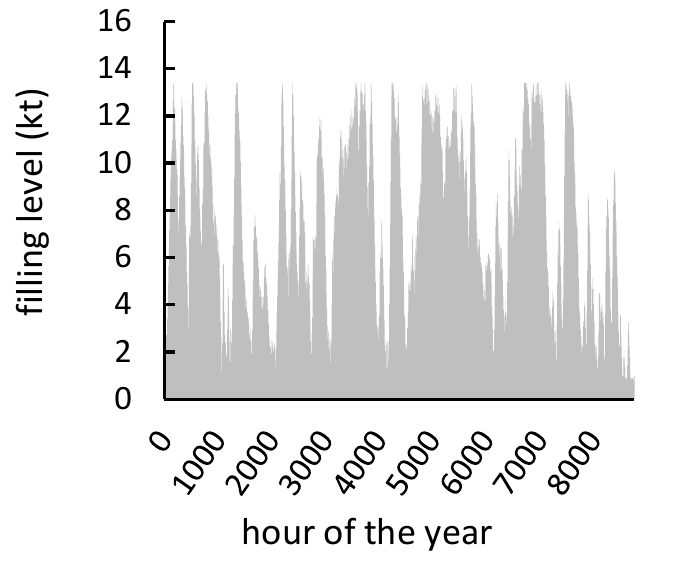}}~~~\subcaptionbox{\label{fig: Utilization of mass storage - LH2}LH$_{2}$ storage}{\includegraphics[width=4.3cm]{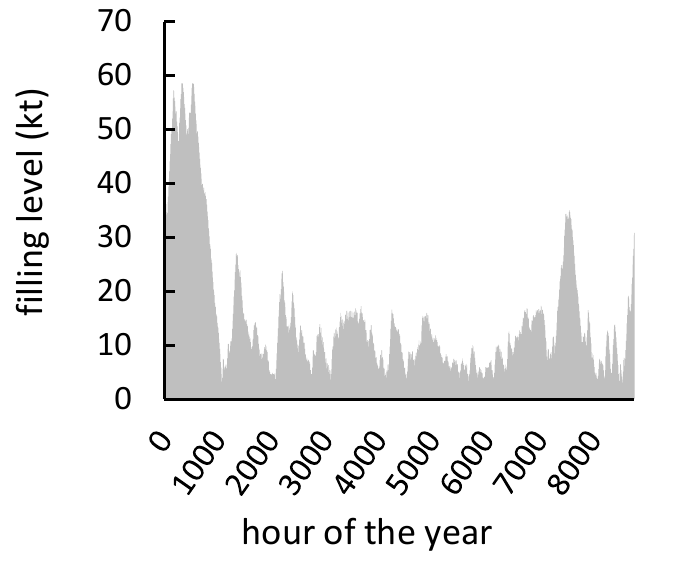}}~~~\subcaptionbox{\label{fig: Utilization of mass storage - LOHC}LOHC storage}{\includegraphics[width=4.3cm]{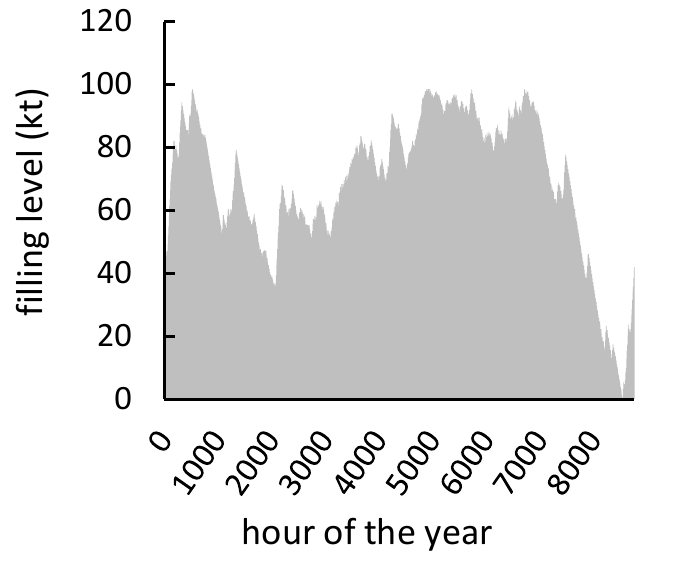}}
\par\end{centering}
\centering{}\caption{\label{fig: Utilization of mass storage}Temporal use pattern of production-site storage in scenario~$Res80$-$Dem25$}
\end{figure}
\par\end{center}
\vspace{-1.4cm}

\subsection{Power sector outcomes reflect drivers for optimal hydrogen supply chains\label{sub: power sector outcomes}}

Figure~\ref{fig: 12 Panel Graph - Capacity} summarizes power sector capacity impacts for the scenarios. Each bar shows the difference of optimal generation capacities compared to the respective baseline without H$_2$~demand. Generally, overall generation capacity increases with growing hydrogen demand and decreases with growing renewable penetration. A higher renewable share leads to higher renewable surplus generation. Large-scale electrolyzers and storage make use of this surplus that would otherwise be curtailed. In fact, in scenarios~$Res80$-$Dem5$ and~$Res80$-$Dem10$, overall electricity generation capacity hardly increases or even decreases because the additional electricity demand for hydrogen production is covered by renewable electricity that would otherwise not be used.

\begin{center}
\begin{figure}[H]
\begin{centering}
\includegraphics[width=0.80\textwidth]{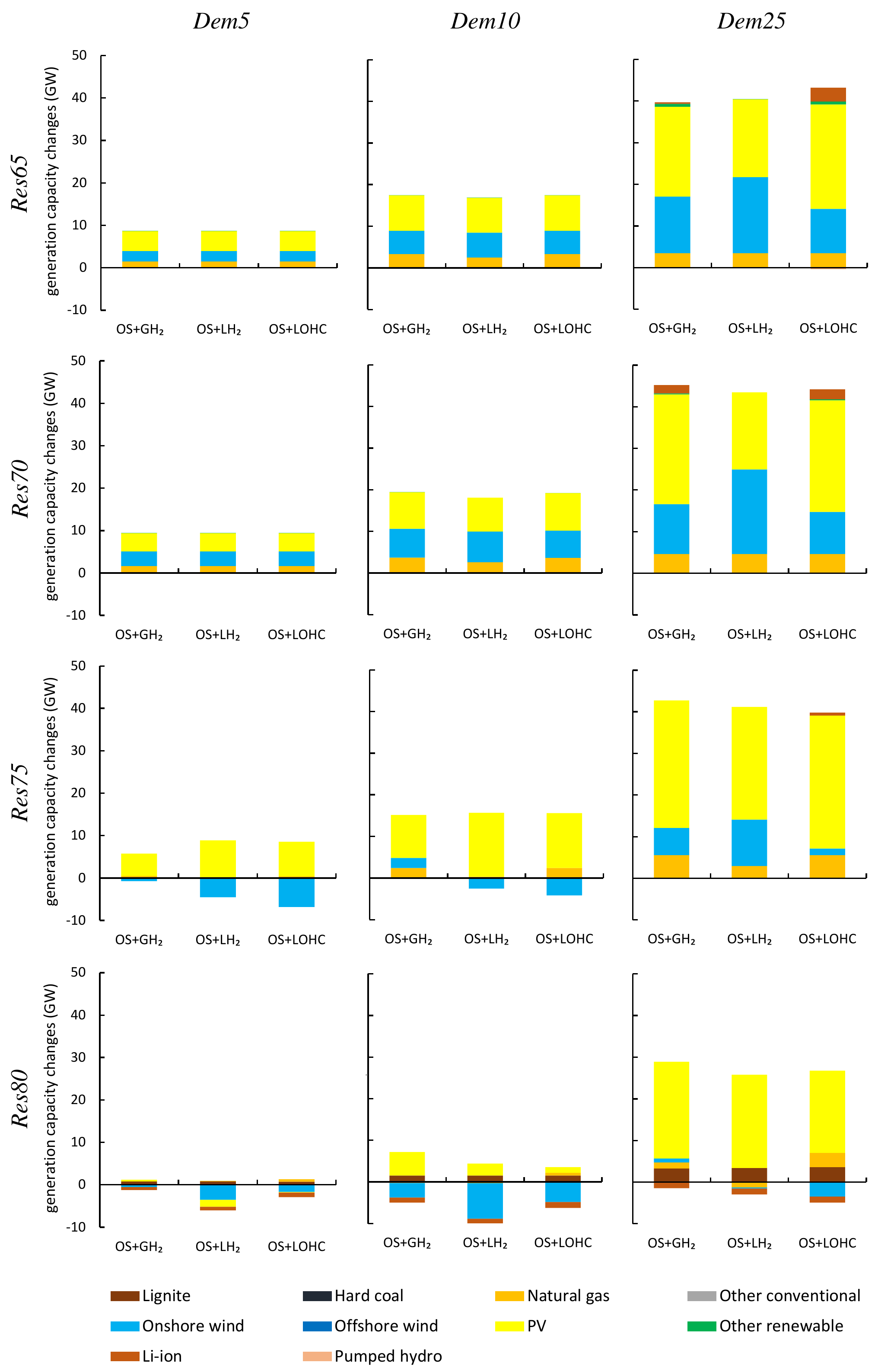} 
\par\end{centering}

\protect\caption{\label{fig: 12 Panel Graph - Capacity}Electricity generation capacity changes compared to the respective baselines without hydrogen for optimal combinations of small-scale and large-scale hydrogen supply chains as shown in Figure~\ref{fig: 12 Panel Graph}.}
\end{figure}
\par\end{center}

Concerning specific technologies, the additional electricity demand for hydrogen supply yields larger optimal solar PV capacities. Additional investments in wind power are lower and the optimal wind power capacity even decreases in some~$Res75$ or~$Res80$ scenarios compared to the respective baseline. Additional wind power would lead to more sustained renewable surplus events, which would be harder to integrate. Offshore wind power is always deployed at the exogenous lower capacity bound of~\unit[17]{GW}. Further, we find a slight increase in the natural gas generation capacity in most scenarios because this is the most economical conventional generation technology to be operated with relatively low full-load hours. Compared to the respective baselines, the supply of hydrogen further tends to increase the optimal electricity storage capacity in the scenarios with lower renewable penetration because temporally inflexible on-site hydrogen production prevails here. In contrast, the optimal electricity storage capacity decreases in the~$Res80$ scenarios. Here, large-scale hydrogen supply chains add a substantial amount of flexibility to the power sector. 

Figure~\ref{fig: 12 Panel Graph - Energy} shows the impact of hydrogen supply chains on yearly energy generation. Across scenarios, wind power is a major source of the additional electricity required for hydrogen supply. Much of this wind power would be curtailed in a power sector without hydrogen. The central driver for this result is that large-scale hydrogen supply chains allow to make better use of variable renewable energy sources, facilitated through longer-term storage. In the~$Res75$ and~$Res80$ scenarios, electricity generation from wind turbines increases substantially although wind capacity barely increases or even decreases (compare Figure~\ref{fig: 12 Panel Graph - Capacity}). Renewable curtailment decreases most in scenario~$Res80$-$Dem25$ with LOHC, where full-load hours of wind power increase by~\unit[19]{\%}. LOHC has the largest capability to integrate renewable surpluses by means of storage and also requires the largest amount of electricity.

Power generation from conventional generators also increases and supplies the part of the additional electricity that is not covered by renewables according to the specified share. In the~$Res65$-$Dem25$ and~$Res70$-$Dem25$ scenarios, with largely inflexible, small-scale electrolysis, this is mainly natural gas-fired power generation. With increasing shares of renewable energy sources, there is a shift to hard coal and lignite. In $Res80$-$Dem25$, the share of lignite in non-renewable power generation is highest. Here, the temporal flexibility of large-scale hydrogen supply chains allows increasing the full-load hours of conventional generation with the highest fixed and lowest variable costs, i.e.,~lignite. Likewise, the use of electricity storage increases compared to the baseline in scenario~$Res65$-$Dem25$, where inflexible small-scale on-site hydrogen supply prevails, but is substituted by large-scale hydrogen flexibility in scenario~$Res80$-$Dem25$.

\begin{center}
\begin{figure}[H]
\begin{centering}
\includegraphics[width=0.8\textwidth]{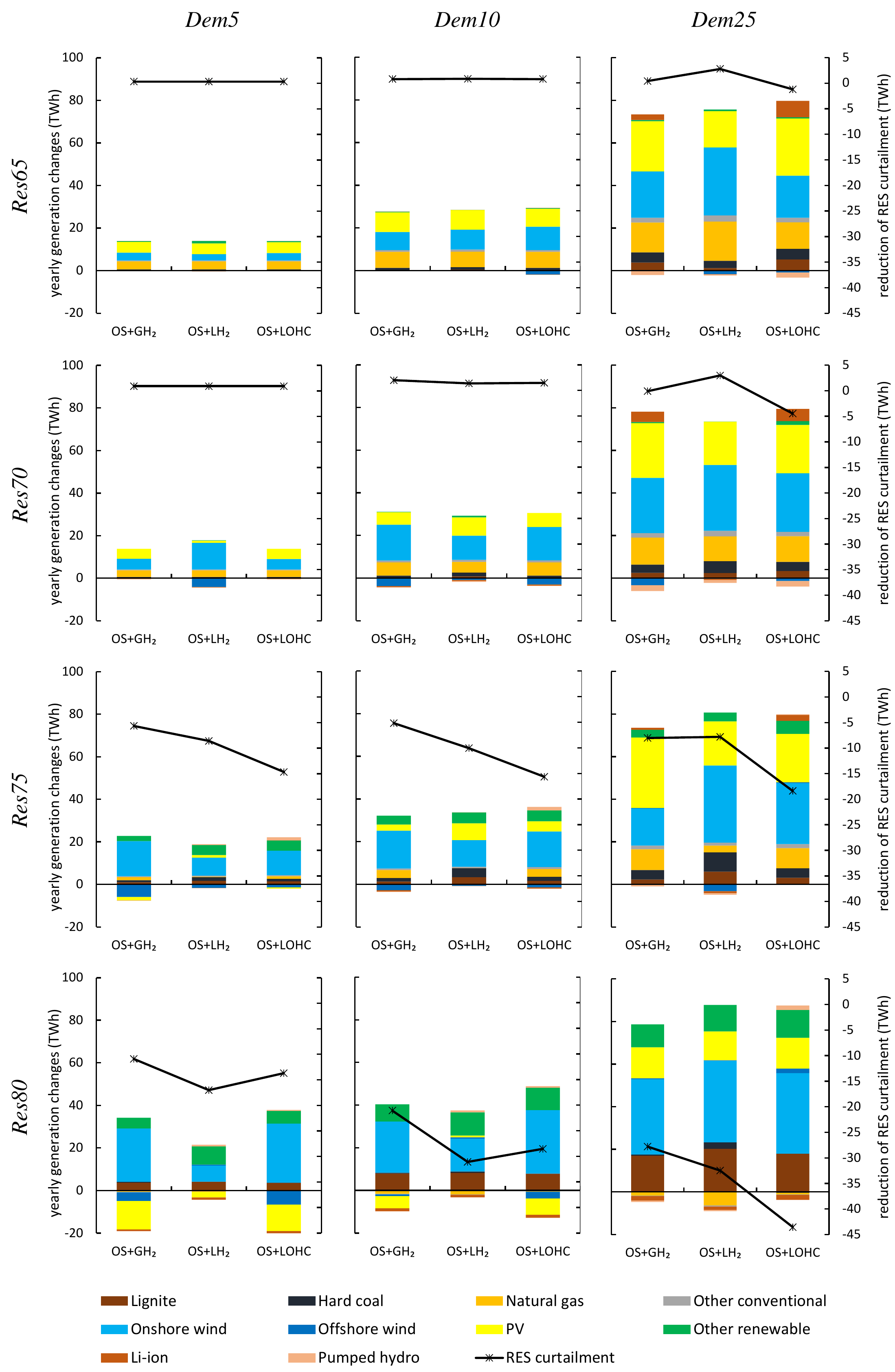} 
\par\end{centering}

\protect\caption{\label{fig: 12 Panel Graph - Energy}Yearly electricity generation changes compared to the respective baselines without hydrogen for optimal combinations of small-scale and large-scale hydrogen supply chains as shown in Figure~\ref{fig: 12 Panel Graph}.}
\end{figure}
\par\end{center}

\newpage

\subsection{\texorpdfstring{CO\textsubscript{2}}{CO₂} emission intensity of hydrogen may not decrease with higher renewable shares}

We calculate the CO$_2$ emission intensity of the hydrogen supplied in two complementary ways (see Section~\ref{sub: Metrics}). The Additional System Emission Intensity of Hydrogen (ASEIH), shown in Figure~\ref{fig: ASEIH}, takes the full power sector effects of hydrogen provision into account. It is defined as the difference of overall CO$_2$ emissions between a scenario with hydrogen and the respective baseline without hydrogen, relative to the total hydrogen demand. The ASEIH mirrors the changes in yearly electricity generation induced by hydrogen supply and ranges between~$6$ and~$13$~kg~CO$_2$ per~kg~H$_2$.

\begin{center}
\begin{figure}[H]
\begin{centering}
\subcaptionbox{\label{fig: ASEIH}Additional System Emission Intensity of Hydrogen (ASEIH)}{\includegraphics[width=1\textwidth]{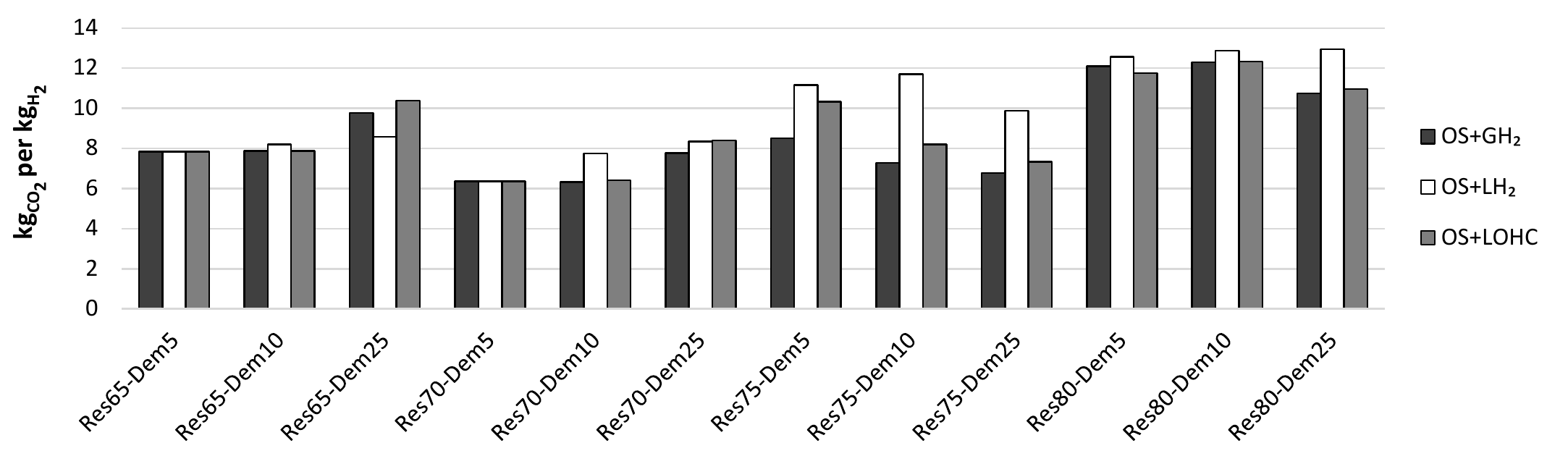}}\newline\subcaptionbox{\label{fig: APEIH}Average Provision Emission Intensity of Hydrogen (APEIH)}{\includegraphics[width=1\textwidth]{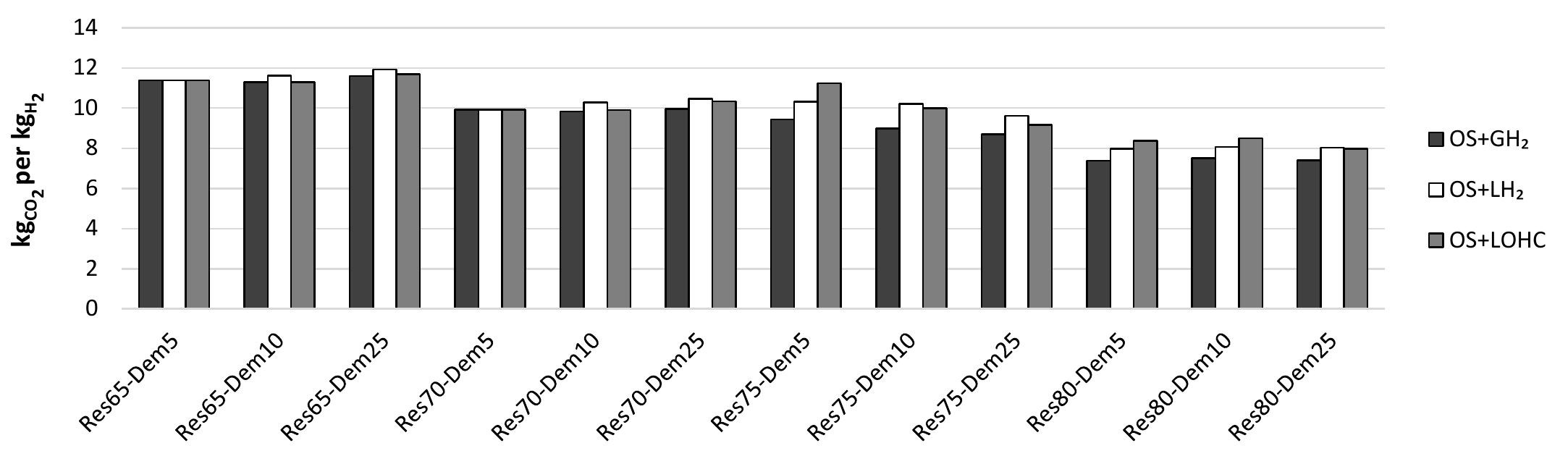}}
\par\end{centering}
\centering{}\caption{\label{fig: system and average co2 emissions}Emission metrics}
\end{figure}
\par\end{center}

Among the~$Res65$ scenarios, the emission intensity of hydrogen is higher for larger hydrogen demand ($Dem25$) because the greater role of flexible large-scale hydrogen infrastructure triggers an increase in coal-fired generation. For a renewable share of~\unit[70]{\%}, the emission intensity is lower because overall power sector emissions decrease and the additional hydrogen demand largely integrates renewables without requiring additional fossil generation. In contrast, for high renewable shares of~\unit[75]{\%} or~\unit[80]{\%}, the ASEIH increases again because the flexibility related to the large-scale hydrogen supply chains allows integrating more coal-fired power generation. This is most pronounced for combinations of small-scale on-site electrolysis and LH$_2$, as the large-scale supply chain has a greater relevance in overall H$_2$ supply compared to~OS+GH$_2$ or~OS+LOHC. Under this metric, thus, the emission intensity of electrolysis-based hydrogen does not necessarily decrease with increasing renewable shares, absent further CO$_2$ regulation.

The second metric, Average Provision Emission Intensity of Hydrogen (APEIH), shown in Figure~\ref{fig: APEIH}, does not capture the differences to an alternative power sector without hydrogen, but is based on CO$_2$ emissions prevailing in the hours of actual hydrogen production. The APEIH ranges between~$7$ and~$12$~kg~CO$_2$ per~kg~H$_2$. The APEIH is highest for the $Res65$~scenarios and generally decreases with increasing renewable shares. It is lowest in supply chains with GH$_2$, slightly higher in with LH$_2$, and highest for LOHC. This largely reflects the differences in energy efficiency among these options.

For lower renewable shares, the APEIH tends to be higher than the ASEIH; for high renewable shares, the APEIH tends to be lower than the ASEIH. That is, a greater renewable penetration decreases the CO$_2$ emissions of the electricity mix used to produce hydrogen (APEIH), but additional emissions induced by H$_2$ do not necessarily decrease (ASEIH). This also indicates that analyses on the emission intensity of green hydrogen should generally be interpreted with care.


\subsection{Power sector benefits of hydrogen\label{sub: benefits}}

We illustrate the power sector benefits of hydrogen supply in two different ways. First, the Average Provision Costs of Hydrogen (APCH) indicate hydrogen costs from a producer perspective. Across all scenarios, the APCH are between around~$5$ and~\unit[8]{€/kg} (Figure~\ref{fig: APCH}). These costs are below the uniform retail price of hydrogen in Germany of around~\unit[9.5]{€/kg} by~2020. In general, the APCH increase with hydrogen demand in all scenarios. With increasing shares of renewable energy, the APCH generally increase slightly, with the exception of scenarios~$Res80$-$Dem5$ and~$Res80$-$Dem10$. Here, supply chain combinations that include LH$_2$ or LOHC lead to lower costs because they can make better use of periods with very low electricity prices, which are frequent in this setting.

In contrast to APCH, the Additional System Costs of Hydrogen (ASCH) metric indicates the costs of hydrogen from a power system perspective. ASCH, which are also shown in Figure~\ref{fig: 12 Panel Graph}, are smaller than APCH in all scenarios. This difference is substantially more pronounced for higher renewable shares (Figure~\ref{fig: diff APCH and system costs}). The ASCH also include the benefits of better renewable energy integration compared to a system without hydrogen. Yet, these benefits cannot be fully internalized by customers at filling stations, as the difference to the more production-oriented APCH metric indicates.

\begin{center}
\begin{figure}[H]
\begin{centering}
\subcaptionbox{\label{fig: APCH}Average Provision Costs of Hydrogen (APCH)}{\includegraphics[width=1\textwidth]{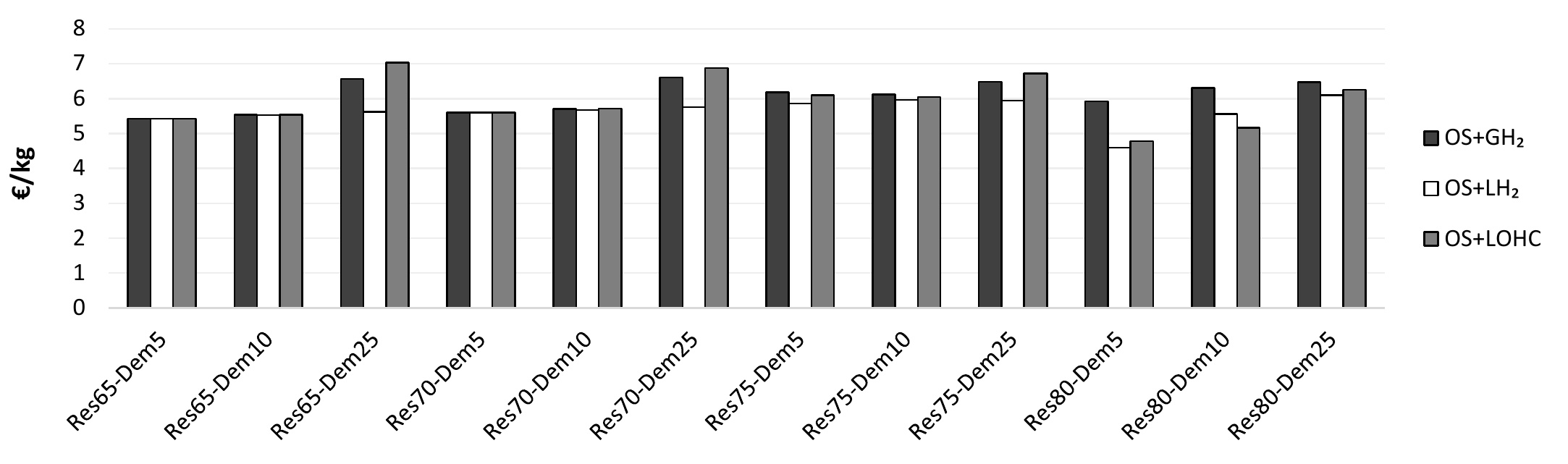}}\newline\subcaptionbox{\label{fig: diff APCH and system costs}Difference between APCH and ASCH}{\includegraphics[width=1\textwidth]{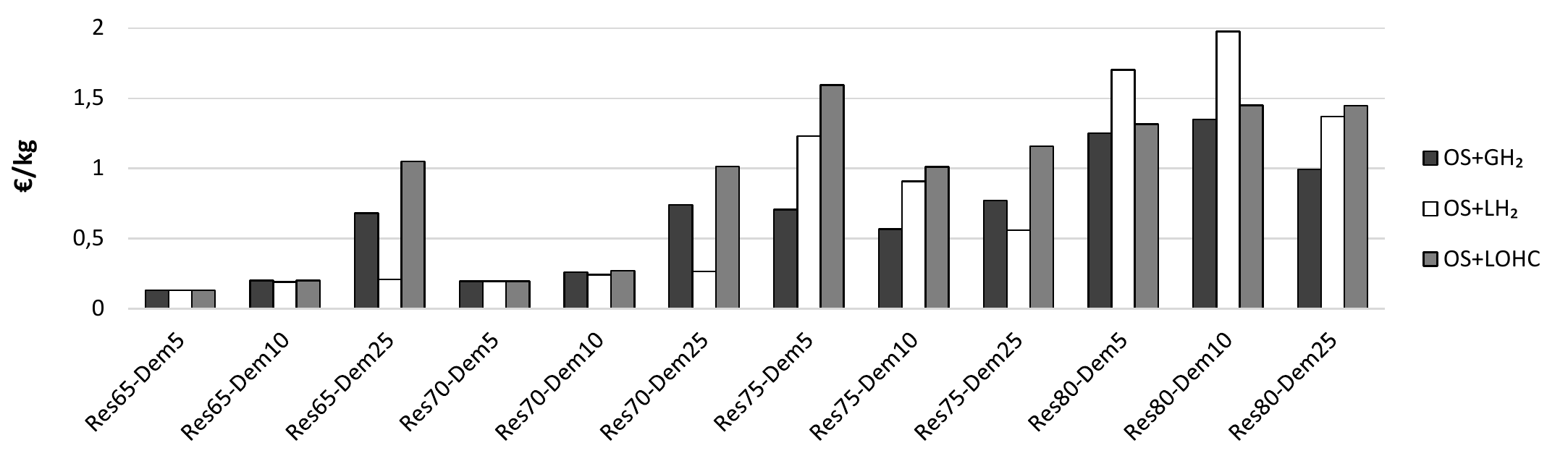}}
\par\end{centering}
\centering{}\caption{\label{fig: two graphs APCH and diff}Average Provision Costs of Hydrogen (APCH) and differences to Additional System Costs of Hydrogen (ASCH).}
\end{figure}
\par\end{center}

\begin{figure}[H]
\centering{}\includegraphics[width=1\textwidth]{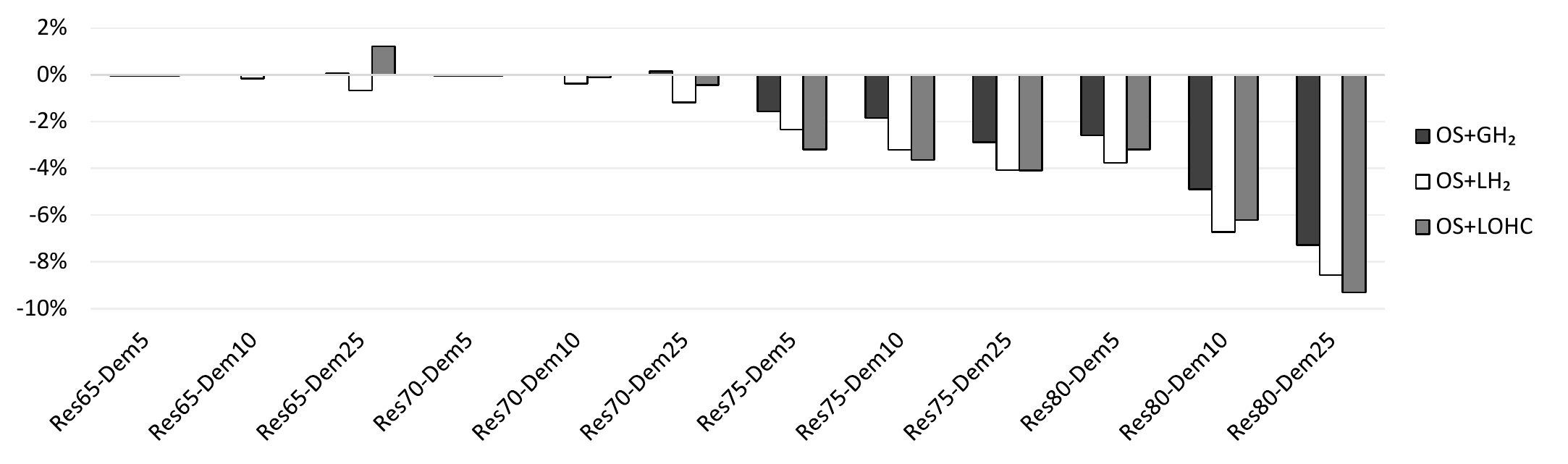}\\
\protect\caption{\label{fig: sce}Effect of hydrogen on System Costs of Electricity (SCE)}
\end{figure}

Second, we illustrate the power sector benefits of different hydrogen supply chains with their impacts on the System Costs of Electricity (SCE, Section~\ref{sub: Metrics}). Here, the total benefits of integrating the power and hydrogen sectors are attributed to the costs of generating electricity. For renewable shares of~\unit[65]{\%} and~\unit[70]{\%}, hydrogen hardly has an impact (Figure~\ref{fig: sce}). Yet, SCE decrease markedly for higher renewable shares, up to more than~\unit[9]{\%} for a combination of small-scale on-site electrolysis and LOHC in the $Res80$-$Dem25$~scenario. The main driver for these benefits, again, is reduced renewable curtailment.

\subsection{Sensitivity analyses: impacts of central parameter assumptions on supply chains\label{sub: sensitivites-main-body}}

Additional model runs show the impact of alternative assumptions for central input parameters (see Section~\ref{sub: Sensitivities}). GH$_2$ and LOHC tend to improve relative to LH$_2$ if the transportation distance decreases, and vice versa, in particular if the share of large-scale production is high. If mass hydrogen storage could be placed at filling stations, this would greatly benefit the small-scale on-site supply chain. GH$_2$ becomes the dominant option for most scenarios if low-cost cavern storage can be developed. LH$_2$ would improve further if boil-off during storage could be avoided. In turn, LOHC would become dominant in most scenarios if free waste heat could be used for dehydrogenation, and if existing transportation and storage infrastructure could be used without additional costs.


\section{Qualitative effects of model limitations\label{sec: limitations}}

Here, we briefly discuss some limitations of the study and how they may qualitatively impact results. Several research design choices we made for clarity and tractability lead to a power sector that is relatively flexibility-constrained. On the demand side, we abstract from a range of potential flexibility sources, such as power-to-heat options, battery-electric vehicles or the use of hydrogen for other purposes than mobility, e.g.,~high-temperature processes in industry. We also abstract from geographical balancing in the European interconnection. Accordingly, we may overestimate renewable surpluses and, in turn, the benefits of flexible hydrogen supply chains that make use of them. We also do not constrain investments in renewable electricity generation in Germany. A cap on renewable capacity deployment, reflecting public acceptance and planning issues, may further increase the relative importance of energy efficiency compared to flexibility.

We further do not consider potential transmission or distribution grid constraints for clarity and generalizability. These can increase the local value of flexible hydrogen supply, particularly in areas with very good renewable energy resources. For example, temporally flexible large-scale hydrogen supply chains may be particularly beneficial in Germany's Northern region, where the best wind power resources are located. 

Likewise, we abstract from hydrogen distribution via pipelines. These could resolve the efficiency-flexibility trade-off, but are likely to be economical only for transporting large amounts of hydrogen between major hubs.


\section{Discussion\label{sec: Discussion-P2H2}}

Our co-optimization of the power and hydrogen sectors highlights that small-scale on-site electrolysis is most beneficial for lower shares of renewable energy sources and low hydrogen demand because energy efficiency matters more than temporal flexibility in such a setting. The power sector benefits of hydrogen are accordingly small. For higher shares of renewables or higher hydrogen demand, large-scale hydrogen infrastructure options gain importance. LH$_2$ provides the best combination of efficiency, flexibility, and investment cost over the majority of scenarios. In particular, temporally flexible large-scale supply chains make use of renewable surplus generation, which allows reducing optimal renewable capacity investment. Yet this flexibility not only facilitates renewable integration in the power sector, but can also increase the use of conventional generation with low marginal costs, with respective carbon emission effects \citep[cp.~][]{Michalski2017a}. The emission intensity of hydrogen thus not necessarily decreases with higher renewable shares, absent further CO$_2$ regulation. 

Overall, the costs of supplying hydrogen at filling stations are relatively similar among optimal supply chain combinations in most of our modeled scenarios. Real-world investment choices would thus have to take additional factors into account that the model analysis cannot not capture. This includes aspects of operational safety and public acceptance, which may favor LOHC, or constraints to renewable energy deployment, which may favor the more energy-efficient options.

While cross-study comparisons are generally challenging due to differences in general model set-ups and parametrizations, our hydrogen cost results are largely in the range of previous analyses. Our Average Provision Costs of Hydrogen (APCH) of~$5$-\unit[8]{€/kg} are relatively similar to the values calculated by \citep{Michalski2017a} and\citep{Rose2020a}, and somewhat at the lower end of the values found by \citep{Reuss2017av} and \citep{Emonts2019a}. Yet we go beyond the previous literature as our research design also allows for differentiating between APCH and Additional System Costs of Hydrogen (ASCH). The latter are generally lower, and increasingly so for higher shares of renewables, as they include the power sector benefits of better renewable integration. In addition to our numerical findings, our presentation and discussion of the APCH and ASCH metrics may be helpful for the energy modeling community. Similar cost calculations could also be carried out for other sector coupling options than green hydrogen. The same is true for the corresponding carbon emission metrics APEIH and ASEIH introduced here.

While our model analysis is parametrized for Germany, our main findings and conclusions should also apply to power sectors in other geographical settings that undergo a transformation from thermal generators toward variable renewable energy sources. Yet, in specific settings where large potentials of dispatchable renewable energy sources are available, such as hydro reservoirs and geothermal energy, the benefits of more flexible hydrogen supply chains are likely to be substantially lower than modeled here. Other countries may also differ from Germany with respect to important hydrogen-related factors such as typical transportation distances, the availability of low-cost hydrogen storage in caverns, the availability of low-cost waste or renewable heat sources for dehydrogenation, and the stock of existing transportation and storage infrastructure. The sensitivities presented in Section~\ref{sub: sensitivites-main-body} qualitatively indicate how our model results depend on these factors.

We conclude that energy system analysts and planners should consider the flexibility and efficiency trade-off of green hydrogen in more detail when assessing its role in future energy transition scenarios. This requires a sufficiently detailed representation of hydrogen supply chains in respective energy modeling tools. To realize flexibility benefits in actual energy markets, policymakers should further redesign tariffs and taxes such that they do not overly distort wholesale price signal along all steps of the hydrogen supply chain~\citep[cf.][]{Guerra2019}, while enabling a fair distribution of the benefits between hydrogen and electricity consumers.

Future research may aim to address some limitations of this study (cf. Section~\ref{sec: limitations}), or explore the efficiency-flexibility trade-off for different hydrogen carriers that allow long-range bulk transport of green hydrogen from remote areas with excellent wind or PV resources, such as Patagonia or Australia. Likewise, extending our analysis to also include the reconversion of hydrogen to electricity in scenarios with full renewable supply would be promising~\citep{Staffell2019, Welder2019}.


\section{Acknowledgments\label{sec: Acknowledgments}}

We thank Markus Reuß and Philipp Runge for fruitful discussions and helpful comments. We are also grateful that Markus Reuß shared a spreadsheet tool to easily calculate electricity demand for compression. We further thank the participants of the following seminars and workshops for valuable feedback: Seminar of the Climate \& Energy College at the University of Melbourne, \unit[100]{\%}~Renewable Energy workshop at the Australian National University, Strommarkttreffen Berlin, Power-to-X Day at Dechema Frankfurt, BB2 research seminar at ifo Munich, and IAEE International Conference. We further thank Amine Sehli, Seyed Saeed Hosseinioun, and Justin Werdin for research assistance. Wolf-Peter Schill carried out parts of this work during a research stay at the Energy Transition Hub at the University of Melbourne. We gratefully acknowledge research funding by the German Federal Ministry of Education and Research via the Kopernikus P2X project, research grant 03SFK2B1.


\section{Author contributions\label{sec: Contributions}}

Conceptualization, W.P.S.; Methodology, F.S., W.P.S., and A.Z.; Software, F.S.; Writing - original draft, F.S., W.P.S, and A.Z.; Writing - Review and Editing, F.S. and W.P.S.; Visualization, F.S. and A.Z.; Project administration and funding acquisition, W.P.S.


\section{Declaration of interests\label{sec: declaration}}
The authors declare no competing interests.

\addcontentsline{toc}{section}{\refname}
\putbib[references-repository-p2x]
\end{bibunit}

\begin{bibunit}[IEEEtranSN]

\renewcommand{\thesection}{SI}

\global\long\def\thefigure{SI.\arabic{figure}}
\unit
\setcounter{figure}{0}

\newpage

\setcounter{page}{1}

\section{Supplementary Information\label{sec: suppl-info}}


\subsection{Cost and emissions metrics\label{sub: Metrics}}

~~~~~~~\textbf{System Costs of Electricity (SCE)} are the total power sector costs related to overall electricity generation. They include all investment, fixed, and variable power sector costs, but exclude the investment, fixed, and (non-electricity) variable costs of the hydrogen supply chains. Using the SCE, the benefits of integrating the power and hydrogen sectors are completely attributed to electricity generation. The SCE treat all electricity generation equally, irrespective of later consumption for conventional electricity demand, demand for hydrogen production and distribution, or losses in the transformation process.

\textbf{Additional System Costs of Hydrogen (ASCH)} are defined as the difference in total system costs between a scenario that includes hydrogen and the respective baseline without hydrogen demand, related to total hydrogen supply. The ASCH factor in the total power sector benefits of hydrogen supply. ASCH are not directly observable for market participants, but relevant from an energy sector planning perspective. 

\textbf{Average Provision Costs of Hydrogen (APCH)}, in contrast, sum the annualized costs of the hydrogen infrastructure and yearly electricity costs for hydrogen production, related to total hydrogen supply. Yearly electricity costs are the product of the hourly shadow prices of the model's energy balance and the hourly electricity demand along the hydrogen supply chain, summed up over all hours of a year. The APCH reflect a producer perspective (excluding taxes and fees that are potentially relevant in real-world settings). For alternative levelized costs of hydrogen (LCOH) concepts, see~\cite{Kuckshinrichs2018}.

The \textbf{Additional System Emission Intensity of Hydrogen (ASEIH)} relates the overall difference of CO$_2$ emissions between a scenario with hydrogen and the respective baseline without hydrogen to the total hydrogen supply. Analogously to the ASCH, this metric takes the full power sector effects of hydrogen provision into account. Like ASCH, ASEIH are not directly observable in an actual market, but relevant from an energy sector planning perspective.

The alternative \textbf{Average Provision Emission Intensity of Hydrogen (APEIH)} metric is calculated by multiplying hourly average emission intensities of electricity generation with respective hourly electricity consumption for hydrogen supply at all steps of the supply chain (including compression, dehydrogenation etc.) and relating this to overall hydrogen provision. Analogously to the APCH, the APEIH assume a producer perspective.


\subsection{Sensitivities\label{sub: Sensitivities}}

We carry out a range of sensitivity calculations to explore how key parameter assumptions affect results. We investigate the effects of varying transportation distances, alternatively assuming that mass storage for small-scale on-site hydrogen supply is available, alternatively assuming that low-cost cavern storage for GH$_{2}$ is available as well as LH$_{2}$ storage without boil-off, and examine cost-free supply of heat as well as of transportation and storage infrastructure for LOHC.


\newpage
\subsubsection{Transportation distance\label{sub: Sensitivity - Transportation}}

Our baseline assumption for the transportation distance of hydrogen produced in large-scale facilities is~\unit[250]{km}, a value which we derive from previous analyses of the German case \citep{Reuss2017a,Runge2019a}. Here, we examine the effects alternative transportation distance assumptions of~$100$ and~\unit[400]{km}. In general, a shorter/longer transportation distance increases/decreases the shares of large-scale hydrogen supply chains in the optimal solution, see Figures~\ref{fig: 12 Panel Graph sensitivity dist-} and~\ref{fig: 12 Panel Graph sensitivity dist+}. Moreover, with a shorter transportation distance, large-scale technologies are now part of the optimal technology portfolio in some scenarios, while for a longer transportation distance, large-scale supply chains drop out in some scenarios.

\begin{center}
\begin{figure}[H]
\begin{centering}
\includegraphics[width=0.8\textwidth]{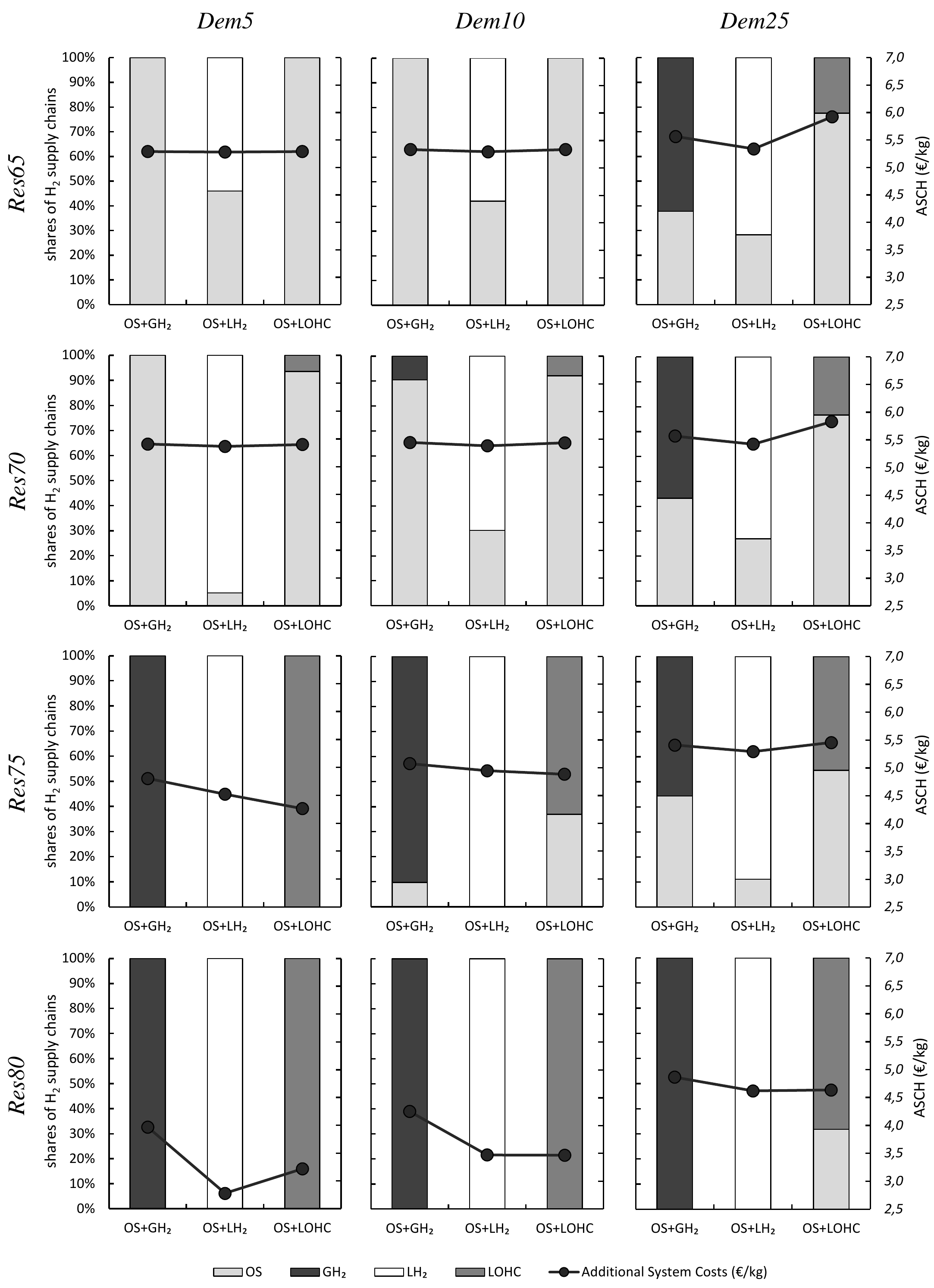} 
\par\end{centering}
\protect\caption{\label{fig: 12 Panel Graph sensitivity dist-}Optimal combinations of small-scale on-site and large-scale hydrogen supply chains and Additional System Costs of Hydrogen (ASCH) for different scenarios - sensitivity with~\unit[100]{km} transportation distance.}
\end{figure}
\par\end{center}

\begin{center}
\begin{figure}[H]
\begin{centering}
\includegraphics[width=0.8\textwidth]{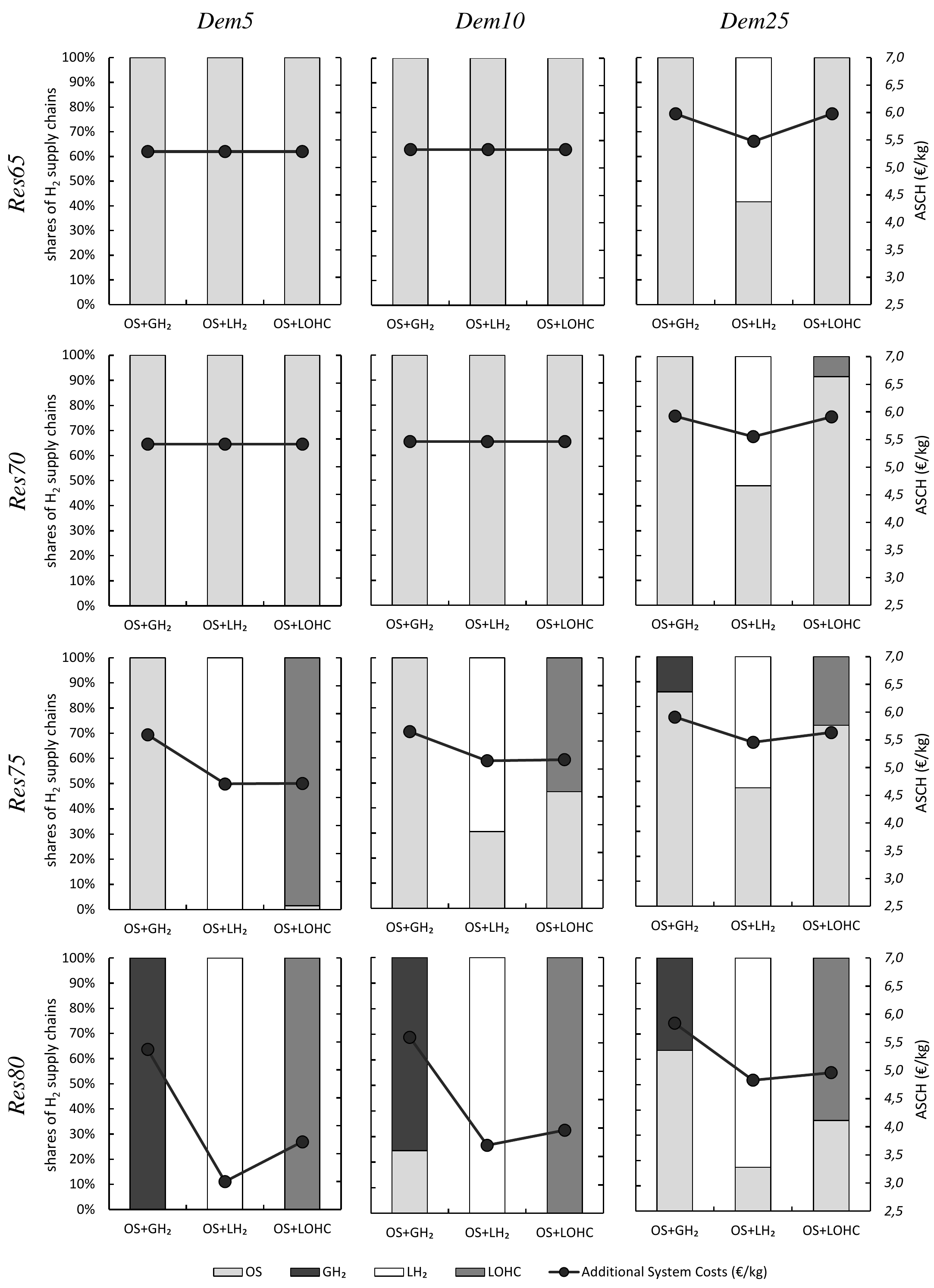} 
\par\end{centering}
\protect\caption{\label{fig: 12 Panel Graph sensitivity dist+}Optimal combinations of small-scale on-site and large-scale hydrogen supply chains and Additional System Costs of Hydrogen (ASCH) for different scenarios - sensitivity with~\unit[400]{km} transportation distance.}
\end{figure}
\par\end{center}

In general, a longer/shorter transportation distance increases/decreases the overall costs of the large-scale hydrogen supply chain. The spread in costs across supply chain combinations within scenarios tends to increase with transportation distance. Yet, the overall least-cost options are robust, with LH$_2$ as dominant large-scale supply chain in the optimal solution. Cost outcomes are fairly robust with respect to the transportation distance because the share of transportation-related costs in the overall costs of hydrogen provision are relatively small.

In more detail, a change in the average transportation distance has two effects on the costs of hydrogen supply. First, variable transportation costs (fuel and driver wage) are proportional to the transportation distance. For the sensitivity calculations with~\unit[400]{km} and~\unit[100]{km} transportation distances, the variable costs increase/decrease by~\unit[60]{\%}. While the relative effect is the same for all three large-scale supply chains, the effect on absolute cost is highest for GH$_2$ and also more pronounced for LOHC than for LH$_2$, see Figure~\ref{fig: Sensitivities- Variable transportation costs}. 

Second, longer/shorter distances imply that each trailer is occupied for a longer/shorter time period. Consequently, the fleet capacity needs to be increased or can be reduced, respectively. Figure~\ref{fig: Sensitivities - Fixed transportation capacity investment costs} shows transportation capacity investment costs per kg of hydrogen supplied through a specific supply chain averaged over all $Res$-$Dem$-scenarios. The pattern is identical to the one for variable costs, yet with less impact in absolute terms.

\begin{center}
\begin{figure}[H]
\begin{centering}
\subcaptionbox{\label{fig: Sensitivities- Variable transportation costs}}{\includegraphics[width=6.5cm]{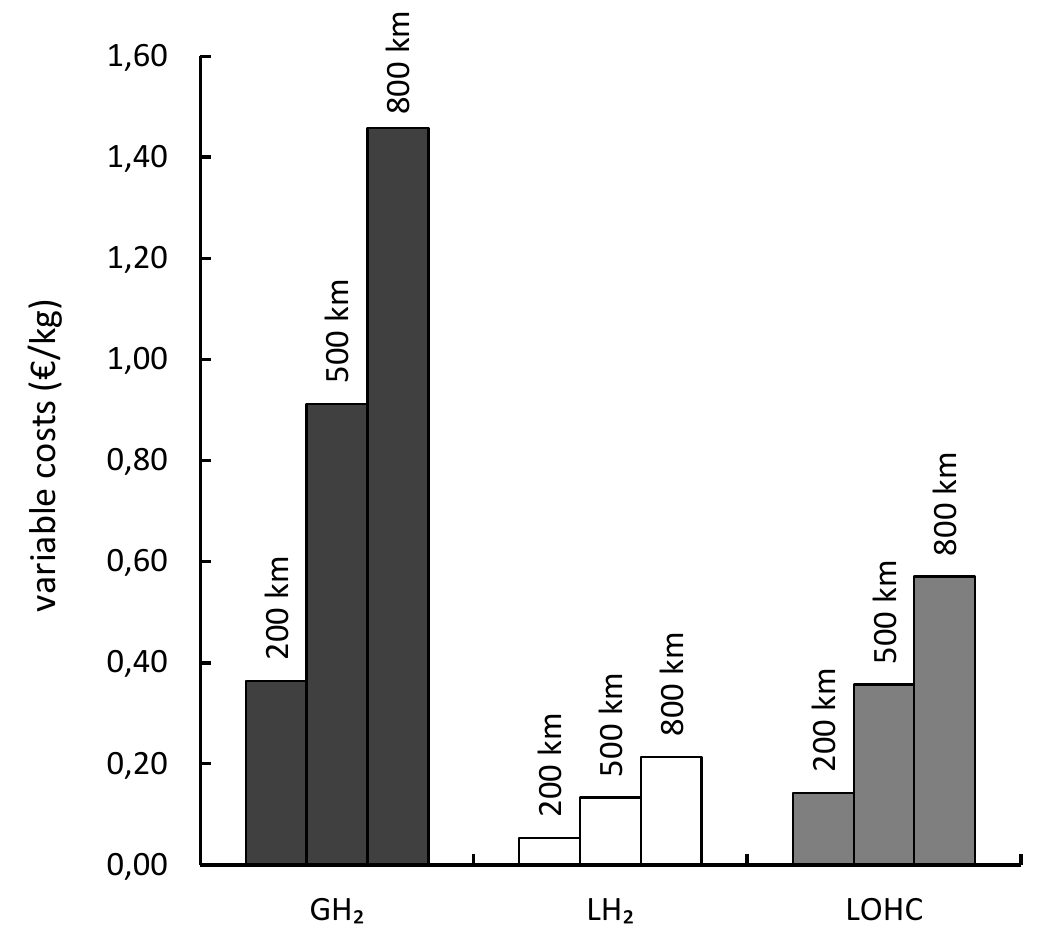}}~~~~~\subcaptionbox{\label{fig: Sensitivities - Fixed transportation capacity investment costs}}{\includegraphics[width=6.5cm]{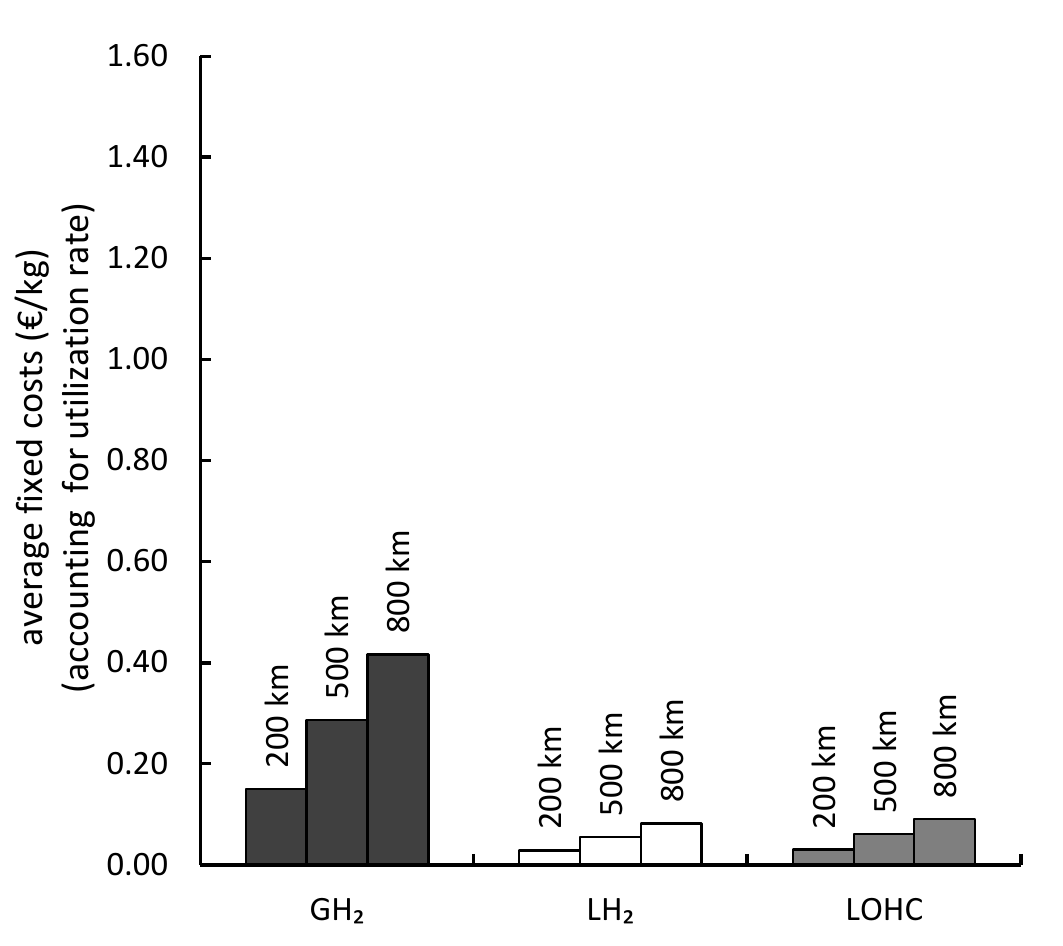}}
\par\end{centering}
\centering{}\caption{\label{fig: Sens - Transportation Costs}Average transportation capacity investment costs and variable costs per~kg of hydrogen supplied through the respective channel.}
\end{figure}
\par\end{center}


\newpage
\subsubsection{Mass storage for small-scale on-site hydrogen supply\label{sub: Mass Storage for Decentralized Production}}

\begin{center}
\begin{figure}[H]
\begin{centering}
\includegraphics[width=0.8\textwidth]{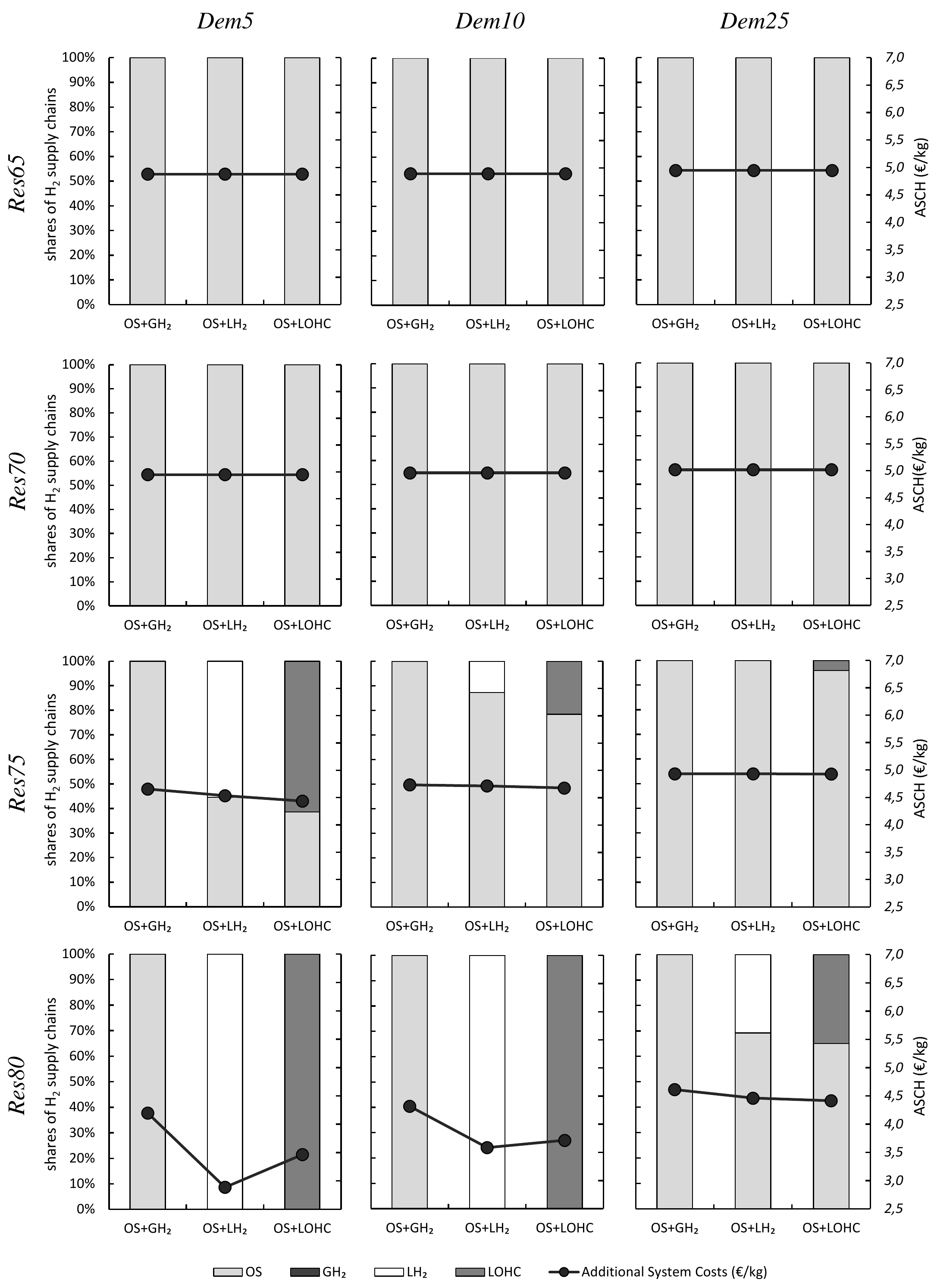} 
\par\end{centering}
\protect\caption{\label{fig: 12 Panel Graph sensitivity flexdec}Optimal combinations of small-scale on-site and large-scale hydrogen supply chains and Additional System Costs of Hydrogen (ASCH) for different scenarios - sensitivity with mass storage available for small-scale on-site production.}
\end{figure}
\par\end{center}

Under baseline assumptions, mass hydrogen storage is not available at filling stations for small-scale supply because of space requirements and security concerns. Alternatively, we assume that relatively cheap mass storage at~\unit[250]{bar} can be deployed at filling stations, with the same techno-economic assumptions as for large-scale GH$_2$ storage. Table~\ref{tab: sensitivity: mass storage for dec} gives an overview of the necessary changes with respect to compression processes and storage infrastructure. 

Consequently, small-scale on-site production of hydrogen becomes more temporally flexible and loses its major disadvantage compared to large-scale production. Given that on-site hydrogen supply t filling stations is more energy-efficient, its share substantially increases for most supply-chain combinations and $Res$-$Dem$-scenarios (Figure~\ref{fig: 12 Panel Graph sensitivity flexdec}), except for those with the highest renewable surpluses, i.e.,~$Res80$-$Dem5$ and~$Res80$-$Dem10$, where all demand is still supplied by large-scale technologies. Here, large-scale production of LH$_{2}$ and LOHC still profits from a larger optimal storage size and the according flexibility. GH$_{2}$ produced in large-scale infrastructures drops out completely. As expected, with the additional flexibility option, the ASCH decrease slightly and the spread in costs between different supply chain combinations within each scenario rather decreases. Finally, the pattern of least-cost options across scenarios is robust, except for scenarios~$Res75$-$Dem25$ and~$Res80$-$Dem25$ where the cost-optimal technology portfolio now contains LOHC rather than LH$_{2}$.


\newpage
\subsubsection{Cavern storage for GH\texorpdfstring{$_{2}$}{₂}\label{sub: Sensitivity - Cavern Storage for GH2}}

\begin{center}
\begin{figure}[H]
\begin{centering}
\includegraphics[width=0.8\textwidth]{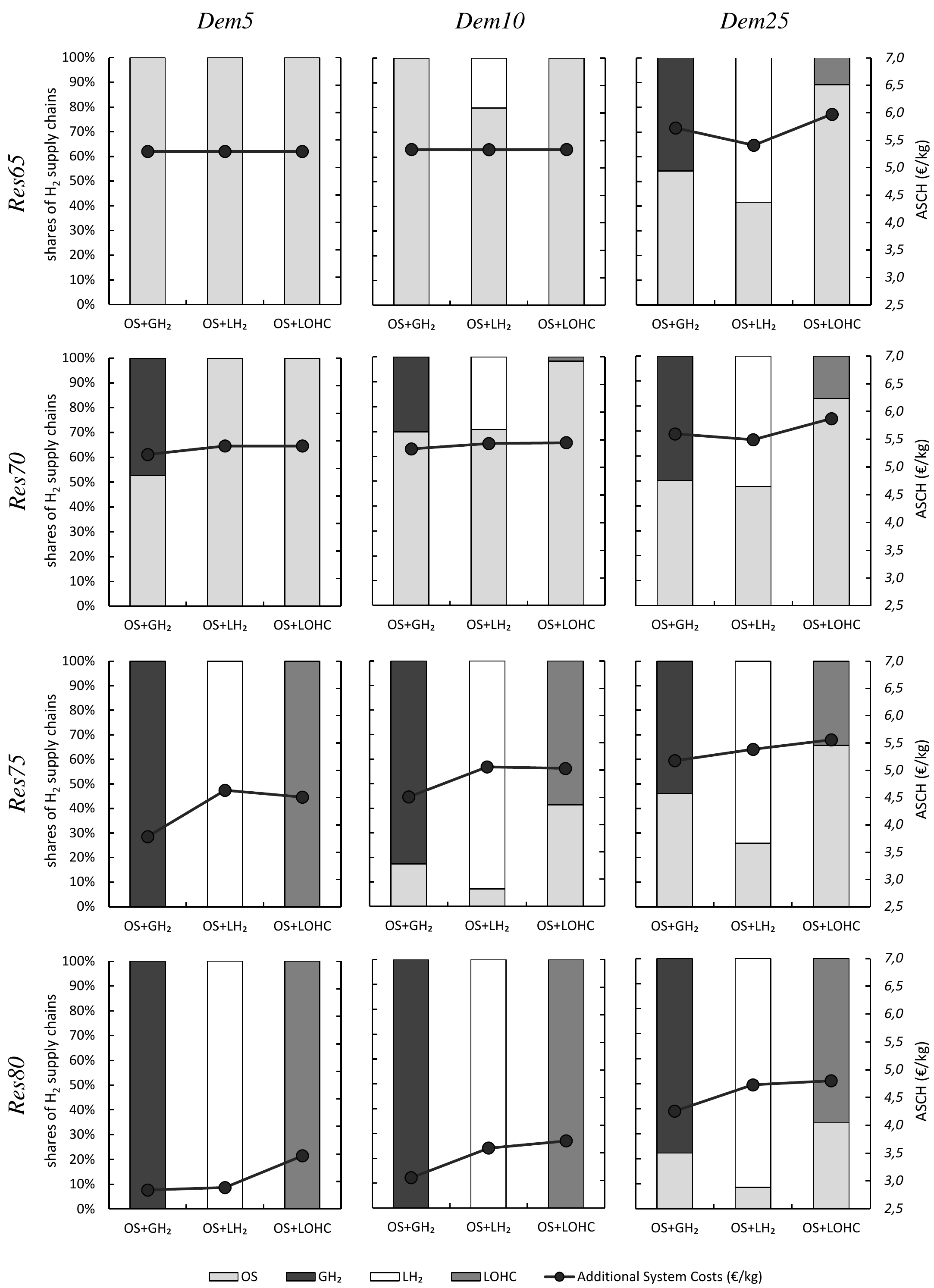} 
\par\end{centering}
\protect\caption{\label{fig: 12 Panel Graph sensitivity cavern}Optimal combinations of small-scale on-site and large-scale hydrogen supply chains and Additional System Costs of Hydrogen (ASCH) for different scenarios - sensitivity with cavern storage available for large-scale GH$_{2}$ production.}
\end{figure}
\par\end{center}

Low-cost cavern storage would provide flexibility for large-scale GH$_2$ production at very low costs of~\unit[3.5]{€/kg}, which is about one third of the costs of LOHC or LH$_2$ storage. Tables~\ref{tab: production site auxiliaries} and~\ref{tab: transporation auxiliaries (before)} list the altered requirements for compression processes. 

If cavern storage is available, the share of large-scale GH$_2$ production increases substantially for all scenarios, see Figure~\ref{fig: 12 Panel Graph sensitivity cavern}. In contrast to the results under default assumptions, the ASCH of the supply chain (DEC+)GH$_2$ are now lower than for the other options in most scenarios, especially if the share of renewable energy sources is high or H$_2$ demand is low. Moreover, Figure~\ref{fig: cav-sto} illustrates that the use of cavern storage exhibits a seasonal pattern, as prevalent for LOHC in the baseline specification, yet with higher storage capacity due to low investment costs. Accordingly, the (non-)availability of cavern storage is a relevant driver of numerical model results.

\begin{center}
\begin{figure}[H]
\begin{centering}
\includegraphics[width=6.75cm]{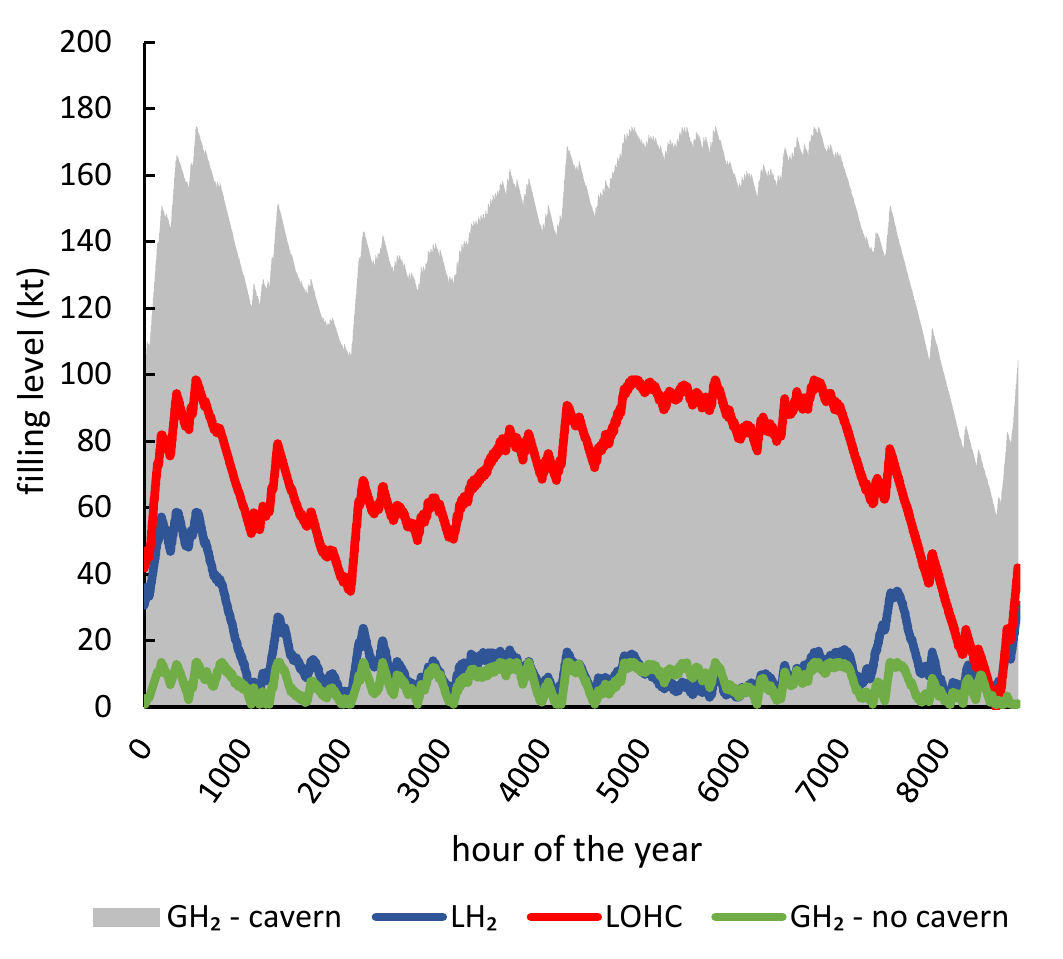} 
\par\end{centering}

\protect\caption{\label{fig: cav-sto}Temporal storage use patterns also including cavern storage for scenario~$Res80$-$Dem25$}
\end{figure}

\par\end{center}


\subsubsection{No boil-off for LH\texorpdfstring{$_{2}$}{₂}\label{sub: Sensitivity - No boil-off for LH2}}

\begin{center}
\begin{figure}[H]
\begin{centering}
\includegraphics[width=0.8\textwidth]{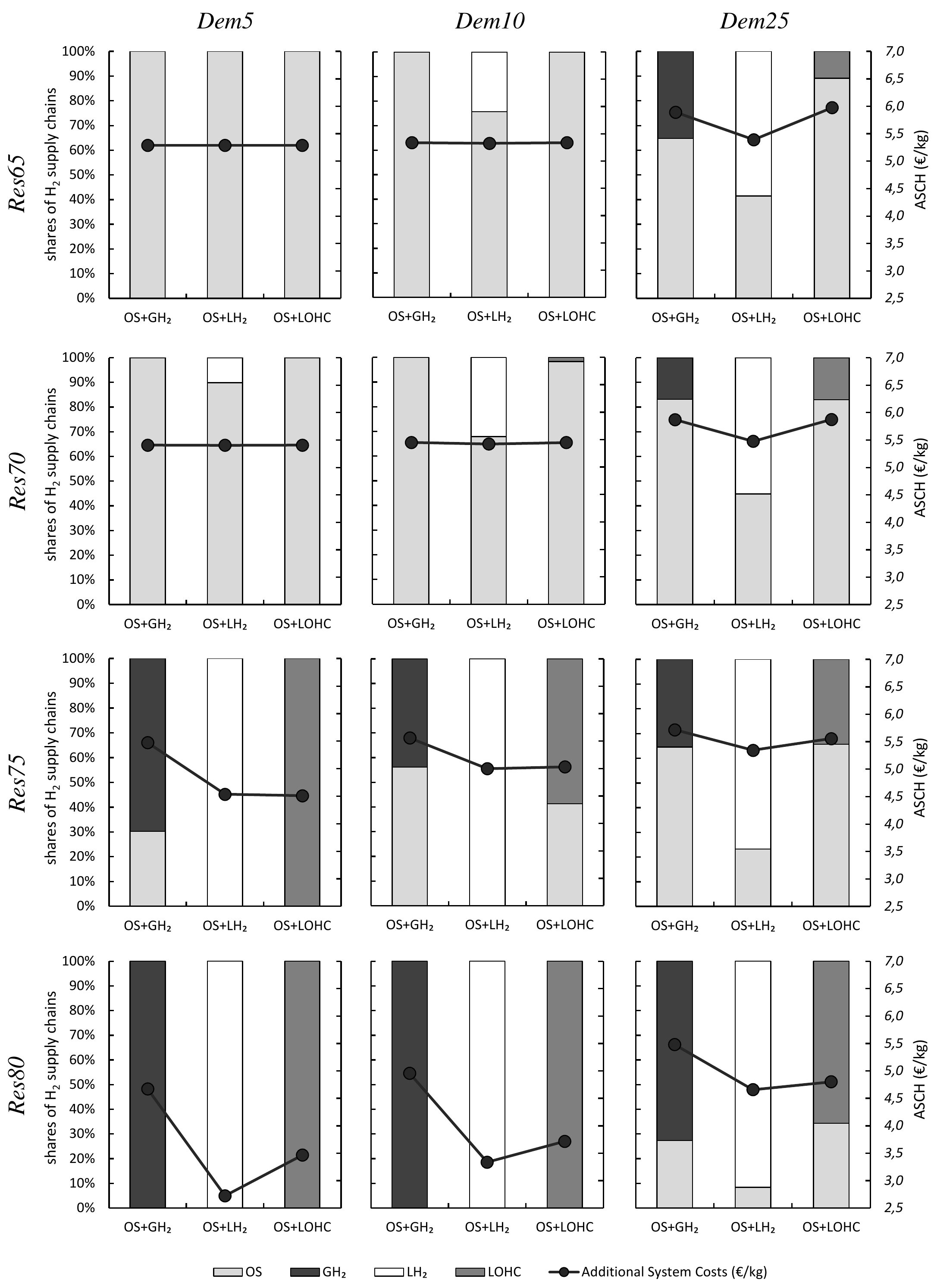} 
\par\end{centering}
\protect\caption{\label{fig: 12 Panel Graph sensitivity - no boil off}Optimal combinations of small-scale on-site and large-scale hydrogen supply chains and Additional System Costs of Hydrogen (ASCH) for different scenarios - sensitivity with no boil-off for LH$_{2}$ storage.}
\end{figure}
\par\end{center}

We assess the effects of LH$_2$ boil-off during storage and transportation by counter-factually setting it to zero. Figure~\ref{fig: 12 Panel Graph sensitivity - no boil off} shows the results. The optimal shares of LH$_2$ compared to on-site hydrogen production at filling stations slightly increase in some cases, but effects are small. The average increase is~$3.2$ percentage points, and the largest increase is~$10.2$ percentage points in scenario~$Res70$-$Dem5$. Likewise, the effect on H$_{2}$ costs is small, with an average cost reduction of~\unit[1.8]{\%} and a maximum decrease of~\unit[7.0]{\%} in scenario~$Res80$-$Dem10$. The pattern of least-cost options is robust with the combination containing LH$_{2}$ now additionally optimal for $Res75$-$Dem10$.

While the effect on costs and optimal technology shares is limited, LH$_2$  without boil-off is better suited as long-term or seasonal storage. Its use pattern changes substantially and resembles that of LOHC under default assumptions. Figure~\ref{fig: Sens - no boil-off} exemplarily illustrates this point for scenario~$Res80$-$Dem25$.

\begin{center}
\begin{figure}[H]
\begin{centering}
\subcaptionbox{\label{fig: Sensitivities - with boil-off}With Boil-Off}{\includegraphics[width=6.5cm]{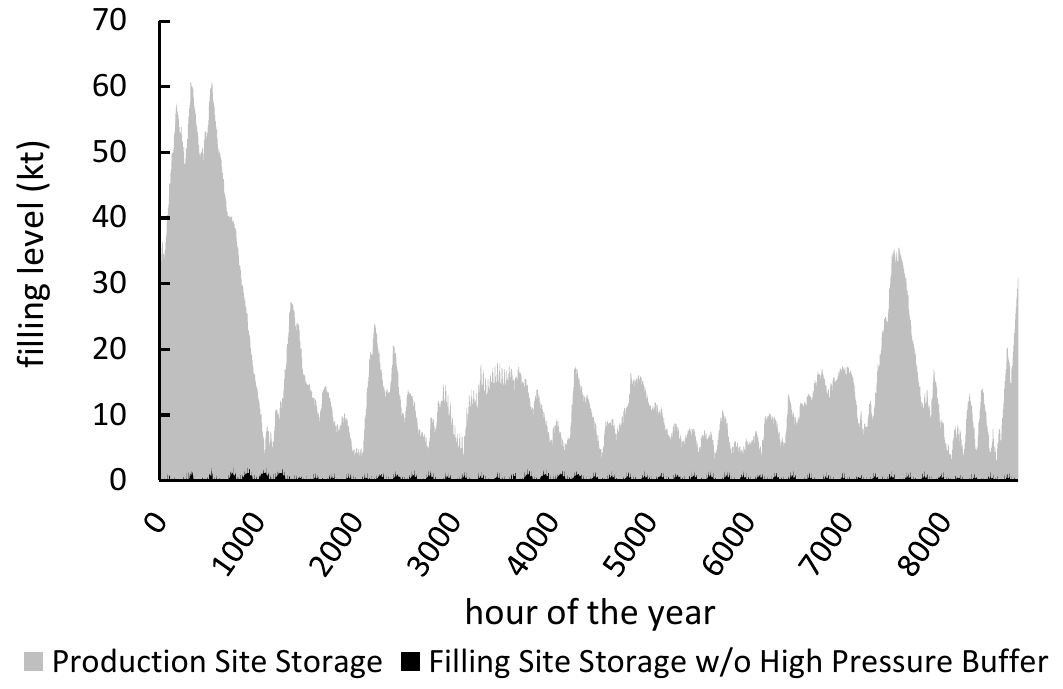}}~~~~~~\subcaptionbox{\label{fig: Sensitivitities - no boil-off}Without Boil-Off}{\includegraphics[width=6.5cm]{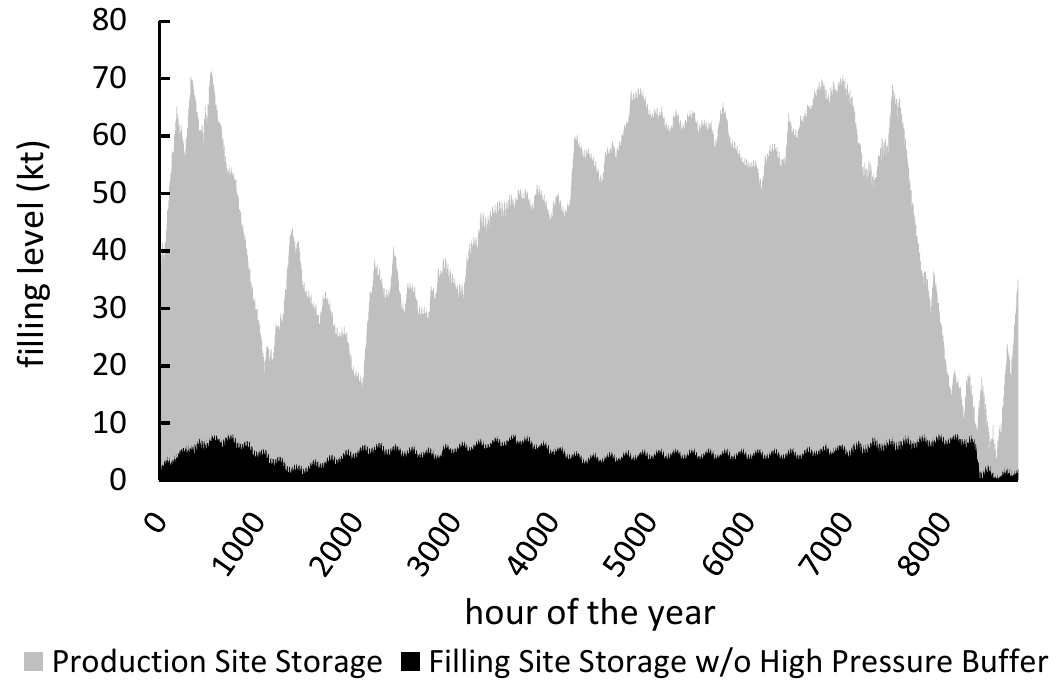}}
\par\end{centering}
\centering{}\caption{\label{fig: Sens - no boil-off}Temporal storage use patterns of LH$_2$ mass storage at the production site for scenario~$Res80$-$Dem25$}
\end{figure}
\par\end{center}

Additionally, we find that LH$_2$ storage at the filling station becomes relatively more important if there is no boil-off. Under default assumptions, boil-off at the filling station was slightly higher than at the production site. Without boil-off, the two storage options are identical in terms of losses over time. Thus, the division of storage between the production and filling sites allows for a more efficient use of transportation capacities. This results in a decrease of transportation infrastructure costs of~\unit[5.5]{\%} per kg of hydrogen in the scenario~$Res80$-$Dem25$.


\newpage
\subsubsection{Free heat supply for LOHC dehydrogenation\label{sub: Sensitivity - Free Heat Supply for Dehydrogenation}}

\begin{center}
\begin{figure}[H]
\begin{centering}
\includegraphics[width=0.8\textwidth]{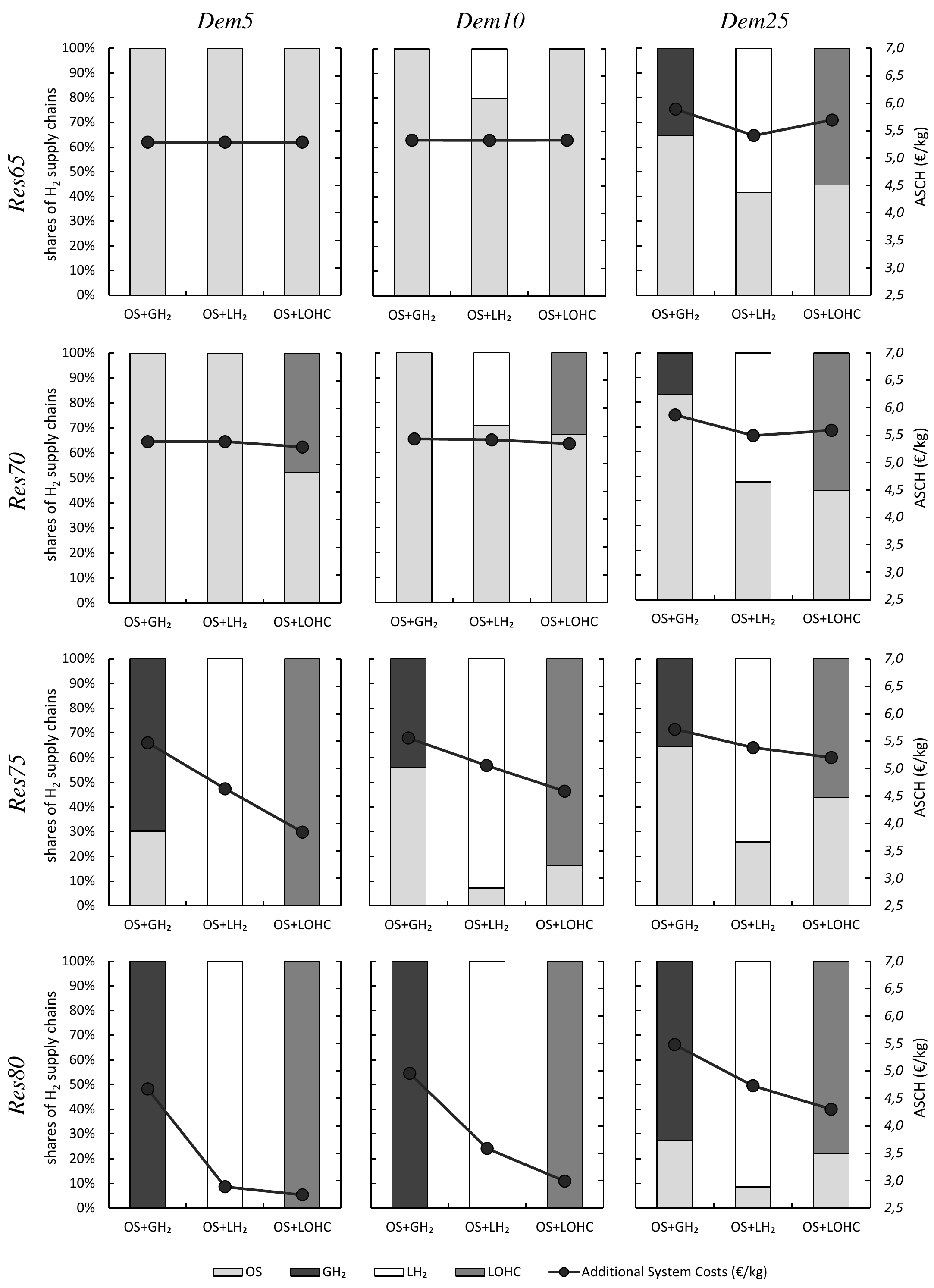} 
\par\end{centering}
\protect\caption{\label{fig: 12 Panel Graph sensitivity free heat}Optimal combinations of small-scale on-site and large-scale hydrogen supply chains and Additional System Costs of Hydrogen (ASCH) for different scenarios - sensitivity with free heat supply for dehydrogenation.}
\end{figure}
\par\end{center}

LOHC has a relatively high electricity demand for dehydrogenation, which is additionally temporally inflexible, that may hold back its extended use. We carry out a sensitivity calculation where the required heat is available free of costs, for instance, because industrial waste heat is available. Figure~\ref{fig: 12 Panel Graph sensitivity free heat} shows the results. Compared to default assumptions, the share of LOHC increases in most scenarios. Also the ASCH for combinations of small-scale on-site electrolysis at filling stations and LOHC decrease. With free heat supply, the LOHC supply chain is the least-cost solution for all scenarios with renewable shares of~\unit[75]{\%} or~\unit[80]{\%}.


\newpage
\subsubsection{Free transportation and production-site storage infrastructure for LOHC\label{sub: Sensitivity - free infrastructure}}

\begin{center}
\begin{figure}[H]
\begin{centering}
\includegraphics[width=0.8\textwidth]{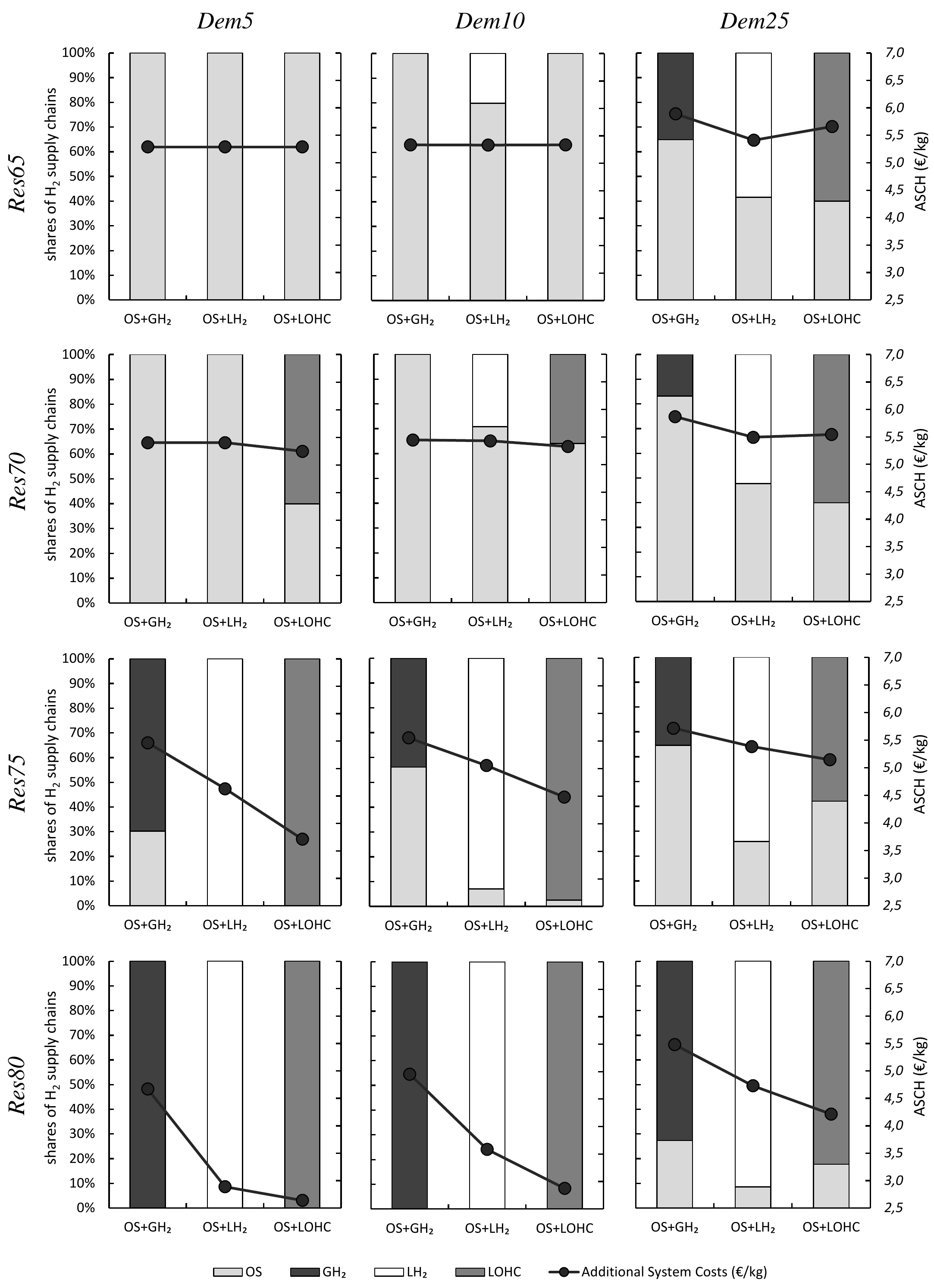} 
\par\end{centering}
\protect\caption{\label{fig: 12 Panel Graph sensitivity free infra}Optimal combinations of small-scale on-site and large-scale hydrogen supply chains and Additional System Costs of Hydrogen (ASCH) for different scenarios - sensitivity with free infrastructure for LOHC storage and transportation.}
\end{figure}
\par\end{center}

Proponents of LOHC argue that existing infrastructure may be used for the LOHC supply chain, especially storage at the production site and filling stations as well as transportation facilities~\citep{Preuster2017}. To address this point in a sensitivity calculation, we assume that storage and transportation capacities do not incur additional costs. Note that the expected lifetime of trailers is~$12$ years. The cost advantage of free transportation capacities would at most last for this time period. The results in Figure~\ref{fig: 12 Panel Graph sensitivity free infra} show that the optimal share of LOHC increases only moderately in many scenarios. In contrast, the ASCH decrease substantially for all supply chains containing LOHC. As for the sensitivity calculation with free heat supply for dehydrogenation, the supply chain involving LOHC is the least-cost option in the scenarios with high renewable penetration also in this case (\unit[75]{\%} or~\unit[80]{\%}).

\newpage
\subsection{Key power sector data\label{sub: power sector Data}}

We apply our model to~2030 scenarios for Germany. To embed the analysis in a plausible mid-term future setting, electricity generation and storage capacities lean on the medium scenario~B of the Grid Development Plan 2019 (\textit{Netzentwicklungsplan},~NEP~\citep{NEP.2018}), an official projection of the German electricity market that transmission system operators base their investments on.

NEP capacities for wind power, both onshore and offshore, solar PV, and battery storage serve as lower bounds for investments. NEP capacities for fossil plants, biomass plants, and run-of-river hydro power serve as upper bounds, where natural gas capacities are split evenly between combined- and open-cycle gas turbines. Coal capacities are largely in line with current German coal phase-out plans that target at most~$9$ and~\unit[8]{GW} lignite and hard coal by~2030, respectively. Investments for pumped storage are bounded from below by today's value and from above by the NEP value. Figure~\ref{fig: apen} summarizes the capacity bounds for the power sector.

\begin{center}
\begin{figure}[h!]
\begin{centering}
\includegraphics[width=1\textwidth]{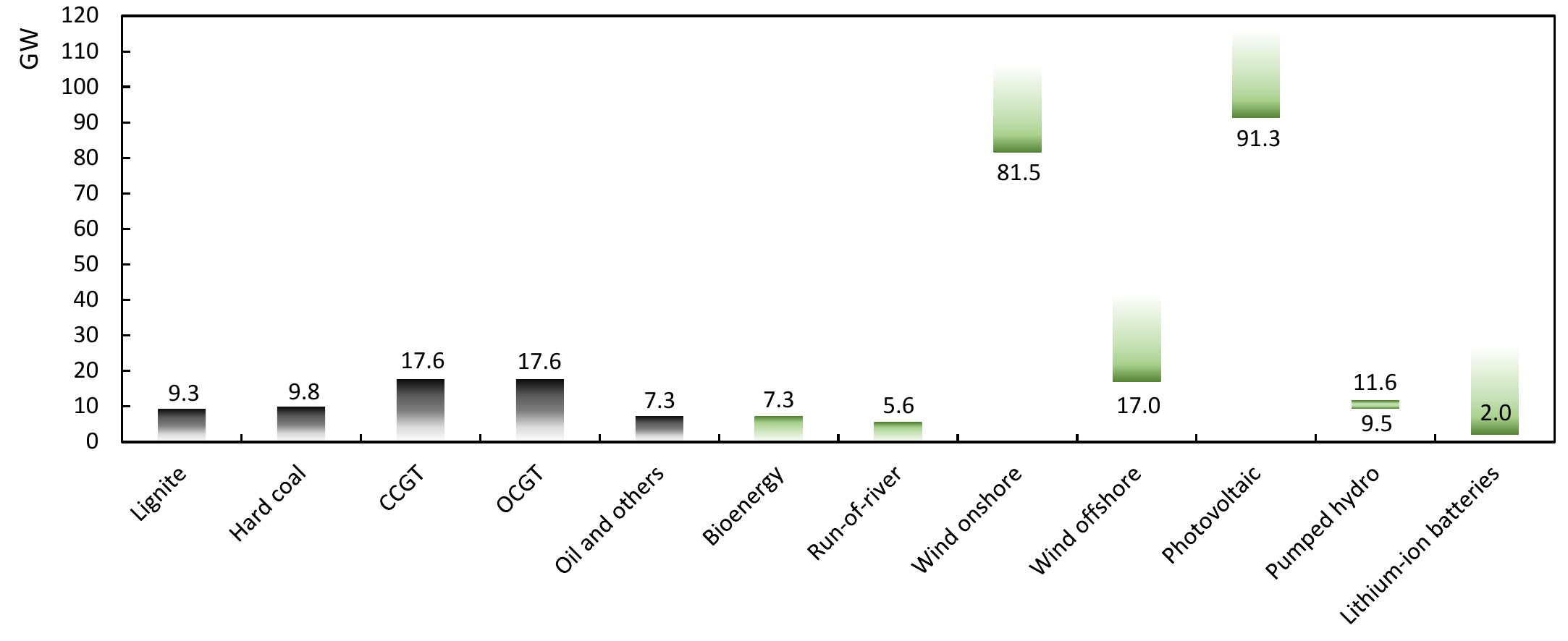} 
\par\end{centering}
\protect\caption{\label{fig: apen} Lower and upper bounds for capacity investments in the power sector}
\end{figure}
\par\end{center}

Cost and technical parameters for power plants~\citep{Schroeder2013} and storage~\citep{Pape2014,Schmidt2017b} are based on established medium-term projections. Fuel costs and the CO$_{2}$ price of~$29.4$\,€/t follow the middle NEP scenario B 2030. The hourly electricity load is representative for an average year and is taken from the Ten-Year Network Development Plan 2030 of the European Network of Transmission System Operators for Electricity~\citep{tyndp.2018}. Annual load sums up to around~$550$~Terawatt hours (TWh). Time series of hourly capacity factors for wind and PV are based on re-analysis data of the average weather year~2012~\citep{Pfenninger2016,Staffell2016}.

All input data is available in a spreadsheet provided together with the open-source model~\citep{Stoeckl2020}.


\newpage
\subsection{Key hydrogen sector data\label{sub: hydrogen sector data}}

In the following, we present key assumptions on the modeled hydrogen sector, including techno-economic parameters of hydrogen infrastructure as well as hydrogen demand. These are central drivers of the results. Full account of all input data is given in Section~\ref{sub: Data-suppl-info}.


\subsubsection{Techno-economic parameters of H\texorpdfstring{$_{2}$}{₂} infrastructure\label{sub: infrastructure}}

PEM electrolysis is six percentage points more efficient than the ALK technology (\unit[71]{\%} versus~\unit[66]{\%}), but has about one-third higher specific investment costs (\unit[905]{€/kW\textsubscript{el}} versus~\unit[688]{€/kW\textsubscript{el}}). Moreover, based on industry data~\citep{Langas2015a}, we assume that investment costs of large-scale electrolysis are~\unit[20]{\%} lower than those of small-scale on-site production at filling stations. 

Cost differences also exist for hydrogen transportation. Trailers for GH$_2$ require high-pressure tubes (\unit[764]{€/kg}), for LH$_2$ an insulated tank (\unit[190]{€/kg}), and for LOHC only a simple standard tank (\unit[93]{€/kg}). Differences in variable costs are determined by the net loading capacity per trailer, where GH$_2$ is most expensive with~\unit[0.91]{€/kg}, compared to~\unit[0.36]{€/kg} and~\unit[0.13]{€/kg} for LOHC and LH$_{2}$, respectively.
%
%
Fuel consumption (Diesel), wages for drivers, and (un-)loading times are assumed to be identical across all supply chains.

Investment costs for hydrogen storage are the central parameter that determines whether flexibility of a supply chain is economical. The costs of GH$_{2}$ storage at~\unit[250]{bar}  (\unit[459]{€/kg}) is substantially higher than for LH$_{2}$ (\unit[14]{€/kg}) and LOHC (\unit[10]{€/kg}). LOHC has a degradation rate of~\unit[0.1]{\%} per supply-cycle, entailing additional costs of~\unit[0.6]{€/kg}. We interpret these costs as LOHC rental rate. High-pressure gaseous (buffer) storage at the filling station is more expensive (\unit[612]{€/kg}) and requires a high minimum filling level in order to ensure pressure above~\unit[700]{bar} for dispensing. This reduces the effective available storage capacity further.

The techno-economic characteristics of the four hydrogen supply chains entail an efficiency-flexibility trade-off with respect to their electricity demand. Small-scale on-site production is relatively energy-efficient but needs to be almost on-time due to a lack of cheap storage options. The three large-scale supply chains are less efficient, but (partly) provide cheap storage options that allow to shift energy-intensive electrolysis to hours with high (renewable) electricity supply. Electricity demand for the remaining, inflexible processes to prepare stored hydrogen for dispensing at the filling station (recompression, cryo-compression, and evaporation or dehydrogenation), is comparably low. Figure~\ref{fig: trade-off} contrasts overall electricity demand with largely inflexible (i.e.,~non-shiftable) electricity demand at the filling station for different hydrogen supply chains across all scenarios. Within-channel deviations (min \& max) are due to the choice of electrolysis technology and losses during storage.

\begin{center}
\begin{figure}[H]
\begin{centering}
\includegraphics[width=6.75cm]{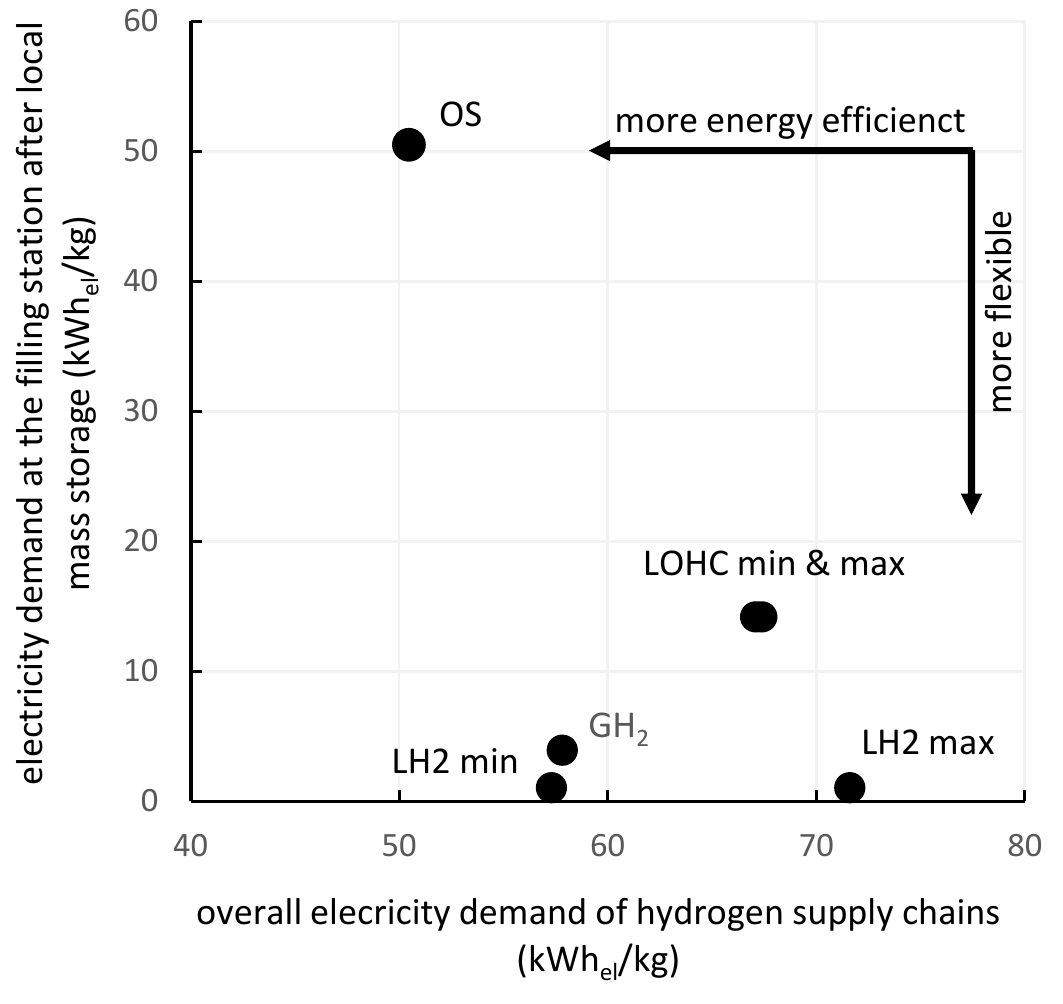} 
\par\end{centering}

\protect\caption{\label{fig: trade-off}The (realized) efficiency-flexibility trade-off for different hydrogen supply chains across all scenarios.}
\end{figure}

\par\end{center}


\subsubsection{H\texorpdfstring{$_{2}$}{₂} demand\label{sub: hydrogen demand data}}

H$_2$ demand for private and public road-based passenger transportation in Germany leans on a forecast for the year~2030~\citep{Schubert2014a}. To convert gasoline and diesel consumption to H$_2$ demand~\citep{Hass2014a}, shares of fuel consumption for~2030 are assumed to be identical to those in~2017~\citep{Radke2017a}. Table~\ref{tab:Traffic Data Projection} shows the resulting demands for the scenarios where~\unit[5]{\%}, \unit[10]{\%}, or~\unit[25]{\%} of private and public road-based passenger traffic in Germany in~2030 is fueled by hydrogen.

The hourly H$_2$ demand profile at the filling stations is assumed to be identical to today's for gasoline and diesel fuel. As data for Germany is not available, we resort to U.S. data for hourly and weekly~\citep{Nexant2008a} as well as for monthly~\citep{US-EIA-2018a} demand characteristics. Moreover, each filling station dispenses at most~\unit[1000]{kg} hydrogen per day~\citep{H2Mobility2010a}. This results in~$976$, $1952$, and~$4880$ filling stations for the~$5$, $10$, and~\unit[25]{\%} demand scenarios, respectively.

\global\long\def\thetable{SI.\arabic{table}}
\unit
\setcounter{table}{0}

\begin{center}
\begin{table}[H]
\protect\caption{\label{tab:Traffic Data Projection}Traffic Data (2030 projection)}

\centering{}%
\begin{tabular}{cr@{\extracolsep{0pt}.}lr@{\extracolsep{0pt}.}l}
\hline 
\noalign{\vskip\doublerulesep}
{\scriptsize{}Scenario } & \multicolumn{4}{c}{{\scriptsize{}H\textsubscript{2} demand}}\tabularnewline
\hline 
\noalign{\vskip\doublerulesep}
{\scriptsize{} } & \multicolumn{2}{c}{{\scriptsize{}TWh}} & \multicolumn{2}{c}{{\scriptsize{}kt}}\tabularnewline
\hline 
\noalign{\vskip\doublerulesep}
{\scriptsize{}\hphantom{2}$\unit[5]{\%}$ } & {\scriptsize{}9}&{\scriptsize{}053} & {\scriptsize{}271}&{\scriptsize{}610}\tabularnewline
{\scriptsize{}$\unit[10]{\%}$ } & {\scriptsize{}18}&{\scriptsize{}160} & {\scriptsize{}543}&{\scriptsize{}220}\tabularnewline
{\scriptsize{}$\unit[25]{\%}$ } & {\scriptsize{}45}&{\scriptsize{}265} & {\scriptsize{}1,358}&{\scriptsize{}050}\tabularnewline
\hline 
\end{tabular}
\end{table}

\par\end{center}

Finally, depending on the average loading capacity and time a car spends at the filling station, a small amount needs to be added to the average costs of hydrogen to cover dispenser costs (around~\unit[0.1]{€} for~\unit[5]{kg} per car with an average filling time of~\unit[7]{min} and a filling station capacity of~\unit[1000]{kg/d}, compare~\citep{Runge2019a}). These costs are identical across all supply chain combinations and, thus, have no effect on their ranking.


\subsubsection{Data tables\label{sub: Data-suppl-info}}

In the following, we list all data and sources for techno-economic parameters concerning the H$_2$ infrastructure. As parameter projections for~2030 are scarce, except for electrolysis, we resort to values for currently existing or planned sites. All cost parameters are stated in euros (\unit{€}). For conversion from U.S.~dollar (\unit{\$}), we assume an exchange rate of one. As the literature on cost parameters does often not provide information on the reference year, we refrain from correcting for inflation. Unless stated otherwise, \unit{kg} is always short for~\unit{kg$_{H_{2}}$}. To calculate electricity demand for compression and scale investment costs, we follow~\citep{Reuss2017a}. Pursuing a conservative approach, we always calculate energy demand for hydrogen compression for the least favorable initial pressure conditions. All data are in terms of the lower heating value (LHV). The costs of water for electrolysis are not taken into account in this analysis as they are negligible in Germany. Finally, OPEX are always stated as \% of CAPEX.
\\

\begin{center}
\begin{table}[H]
\protect\caption{\label{tab: general assumptions}General assumptions}

\begin{centering}
{\scriptsize{}}%
\begin{tabular}{ll}
\hline 
 & {\scriptsize{}Value}\tabularnewline
\hline 
{\scriptsize{}Average transportation distance (one-way) \citep{Reuss2017a}} & {\scriptsize{}250\,km}\tabularnewline
{\scriptsize{}Average transportation speed \citep{Reuss2017a}} & {\scriptsize{}$\unitfrac[50]{km}{h}$}\tabularnewline
{\scriptsize{}Interest rate} & {\scriptsize{}4\,\%}\tabularnewline
{\scriptsize{}Loading (LOHC) \citep{Eypasch2017a}} & {\scriptsize{}6.2\,\textdegree weight-\%}\tabularnewline
{\scriptsize{}LOHC costs\textsuperscript{a} \citep{Teichmann2012a}} & {\scriptsize{}$\unitfrac[4]{\text{€}}{kg_{LOHC}}$}\tabularnewline
\hline 
\end{tabular}
\par\end{centering}{\scriptsize \par}

\begin{spacing}{0.40000000000000002}
\centering{}%
\begin{tabular}{p{9.1cm}}
{\tiny{}a: LOHC has is a degradation rate of $2\times\unit[0.1]{\%}$
(hydrogenation \& dehydrogenation)~\citep{Teichmann2012a} per supply-cycle, entailing additional costs of~\unit[0.13]{€/kg}. We interpret these costs as LOHC rental rate.}\tabularnewline
\end{tabular}\end{spacing}
\end{table}
{\scriptsize{} }
\par\end{center}{\scriptsize \par}


\begin{center}
\begin{table}[H]
\protect\caption{\label{tab: electrolysis cost assumptions}Assumptions for different electrolysis technologies for 2030}

\begin{centering}
\begin{tabular}{lll}
\hline 
 & {\scriptsize{}ALK} & {\scriptsize{}PEM}\tabularnewline
\hline 
{\scriptsize{}CAPEX $\left(\unitfrac{\text{€}}{kW_{el}}\right)$\textsuperscript{a}
\citep{Schmidt2017a,Langas2015a}} & {\scriptsize{}550} & {\scriptsize{}724}\tabularnewline
{\scriptsize{}OPEX (\%) \citep{Bertuccioli2014a}} & {\scriptsize{}1.5} & {\scriptsize{}1.5}\tabularnewline
{\scriptsize{}Depreciation period (a)\textsuperscript{a,\,d} \citep{Schmidt2017a,Bertuccioli2014a}} & {\scriptsize{}10} & {\scriptsize{}10}\tabularnewline
{\scriptsize{}Efficiency (\%)\textsuperscript{c} \citep{Bertuccioli2014a}} & {\scriptsize{}66} & {\scriptsize{}71}\tabularnewline
{\scriptsize{}Pressure out (bar) \citep{Schmidt2017a,Carmo2013a,Bertuccioli2014a}} & {\scriptsize{}30} & {\scriptsize{}30}\tabularnewline
{\scriptsize{}Scale advantage (\%)\textsuperscript{b} \citep{Langas2015a}} & {\scriptsize{}20} & {\scriptsize{}20}\tabularnewline
\hline 
\end{tabular}
\par\end{centering}

\begin{spacing}{0.40000000000000002}
\centering{}%
\begin{tabular}{p{7cm}}
{\tiny{}a: Based on a~$\unit[10]{MW_{el}}$ electrolysis system with~$2$ times the current R\&D investment and production scale-up.}{\tiny \par}

{\tiny{}b: Cost advantage when scaling up from~$\unit[2.2]{MW_{el}}$
to~$\unit[10]{MW_{el}}$. The output of a~$\unit[2.2]{MW_{el}}$ and~$\unit[10]{MW_{el}}$ electrolyzer with an efficiency of~$68.5$\,\%
(the center of our assumptions for ALK and PEM) is equal to~\unit[45]{kg/h} and~\unit[206]{kg/h}, respectively.}{\tiny \par}

{\tiny{}c: At the system level, including power supply, system control,
gas drying (purity at least~$99.4$\,\%). Excluding external compression,
external purification, and hydrogen storage.}{\tiny \par}

{\tiny{}d: $60,000$\,h operation at an utilization rate of~$70$\,\%.}\tabularnewline
\end{tabular}\end{spacing}
\end{table}

\par\end{center}


\begin{center}
\begin{table}[H]
\protect\caption{\label{tab: production site auxiliaries}Assumptions for different storage preparation processes (production site)}

\begin{centering}
\begin{tabular}{llllll}
\hline 
 & {\scriptsize{}GH\textsubscript{{\scriptsize{}2}} (S)} & {\scriptsize{}GH\textsubscript{{\scriptsize{}2}} (L)} & {\scriptsize{}GH\textsubscript{{\scriptsize{}2}}\textsuperscript{{\scriptsize{}cav.}}
(L)} & {\scriptsize{}LH\textsubscript{{\scriptsize{}2}} (L)} & {\scriptsize{}LOHC (L)}\tabularnewline
 &  & {\scriptsize{}\citep{Elgowainy2015a}} & {\scriptsize{}\citep{Elgowainy2015a}} & {\scriptsize{}\citep{Stolzenburg2013a}} & {\scriptsize{}\citep{McClaine2015a,Teichmann2012a,Reuss2017a,Eypasch2017a,Muller2015a}}\tabularnewline
\hline 
{\scriptsize{}Activity} & {\scriptsize{}-} & {\scriptsize{}compression} & {\scriptsize{}compression} & {\scriptsize{}liquefaction} & {\scriptsize{}hydrogenation}\tabularnewline
{\scriptsize{}CAPEX-base (€)} & {\scriptsize{}-} & {\scriptsize{}40,528} & {\scriptsize{}40,528} & {\scriptsize{}643,700} & {\scriptsize{}74,657 \citep{Eypasch2017a}}\tabularnewline
{\scriptsize{}CAPEX-comparison} & {\scriptsize{}-} & {\scriptsize{}$\unit[1]{kW_{el}}$} & {\scriptsize{}$\unit[1]{kW_{el}}$} & {\scriptsize{}1\,kg} & {\scriptsize{}1\,kg}\tabularnewline
{\scriptsize{}Scale} & {\scriptsize{}-} & {\scriptsize{}0.4603} & {\scriptsize{}0.4603} & {\scriptsize{}2/3} & {\scriptsize{}2/3}\tabularnewline
{\scriptsize{}Ref.-Capacity $\left(\unitfrac{kg}{h}\right)$} & {\scriptsize{}-} & {\scriptsize{}206} & {\scriptsize{}206} & {\scriptsize{}1030} & {\scriptsize{}1030}\tabularnewline
{\scriptsize{}CAPEX-scaled $\left(\unitfrac{\text{€}}{kg}\right)$\textsuperscript{a}} & {\scriptsize{}-} & {\scriptsize{}2,923} & {\scriptsize{}2,672} & {\scriptsize{}63,739} & {\scriptsize{}7,392 \citep{Eypasch2017a}}\tabularnewline
{\scriptsize{}OPEX (\%)} & {\scriptsize{}-} & {\scriptsize{}4} & {\scriptsize{}4} & {\scriptsize{}4} & {\scriptsize{}4}\tabularnewline
{\scriptsize{}Depreciation period (a)} & {\scriptsize{}-} & {\scriptsize{}15} & {\scriptsize{}15} & {\scriptsize{}30} & {\scriptsize{}20}\tabularnewline
{\scriptsize{}Pressure in (bar)} & {\scriptsize{}-} & {\scriptsize{}30} & {\scriptsize{}30} & {\scriptsize{}30 (20 nec.)} & {\scriptsize{}30}\tabularnewline
{\scriptsize{}Pressure out (bar)} & {\scriptsize{}-} & {\scriptsize{}250} & {\scriptsize{}180} & {\scriptsize{}2} & {\scriptsize{}-}\tabularnewline
{\scriptsize{}Compression stages} & {\scriptsize{}-} & {\scriptsize{}2} & {\scriptsize{}2} & {\scriptsize{}-} & {\scriptsize{}-}\tabularnewline
{\scriptsize{}Elec. Demand $\left(\unitfrac{kWh}{kg}\right)$} & {\scriptsize{}-} & {\scriptsize{}1.707} & {\scriptsize{}1.402} & {\scriptsize{}6.78} & {\scriptsize{}0.37}\tabularnewline
{\scriptsize{}Heat Demand $\left(\unitfrac{kWh}{kg}\right)$} & {\scriptsize{}-} & {\scriptsize{}-} & {\scriptsize{}-} & {\scriptsize{}-} & {\scriptsize{}-8.9}\tabularnewline
{\scriptsize{}Losses (\%)} & {\scriptsize{}-} & {\scriptsize{}0.5} & {\scriptsize{}0.5} & {\scriptsize{}1.625} & {\scriptsize{}3}\tabularnewline
\hline 
\end{tabular}
\par\end{centering}

\begin{spacing}{0.40000000000000002}
\centering{}%
\begin{tabular}{p{13.9cm}}
\textit{\tiny{}Abbreviations:}{\tiny{} cav.: cavern; (S): small-scale on-site supply chain; (L): large-scale supply chain}{\tiny \par}
{\tiny{}a: For~$\unit[10]{MW_{el}}$ $\left(\unitfrac[206]{kg}{h}\right)$
electrolysis capacity, the maximum daily throughput is almost~$5$\,t
of hydrogen. For non-stacked processes such as liquefaction and hydrogenation, we assume a throughput of~\unit[1030]{kg/h} which would be equal to the hydrogen production of a~$\unit[50]{MW_{el}}$ electrolyzer.}\tabularnewline
\end{tabular}\end{spacing}
\end{table}

\par\end{center}


\begin{center}
\begin{table}[H]
\protect\caption{\label{tab: production site storage}Assumptions for different storage types (production site)}

\begin{centering}
\begin{tabular}{llllll}
\hline 
 & {\scriptsize{}GH\textsubscript{{\scriptsize{}2}} (S)} & {\scriptsize{}GH\textsubscript{{\scriptsize{}2}} (L)} & {\scriptsize{}GH\textsubscript{{\scriptsize{}2}}\textsuperscript{{\scriptsize{}cav.}}
(L)} & {\scriptsize{}LH\textsubscript{{\scriptsize{}2}} (L)} & {\scriptsize{}LOHC (L)}\tabularnewline
 &  & {\scriptsize{}\citep{Parks2014a}} & {\scriptsize{}\citep{Kruck2013a}} & {\scriptsize{}\citep{US-DOE-2015a}} & {\scriptsize{}\citep{Reuss2017a}}\tabularnewline
\hline 
{\scriptsize{}CAPEX-base (€)} & {\scriptsize{}-} & {\scriptsize{}450} & {\scriptsize{}3.5} & {\scriptsize{}13.31} & {\scriptsize{}10}\tabularnewline
{\scriptsize{}CAPEX-comparison} & {\scriptsize{}-} & {\scriptsize{}1\,kg} & {\scriptsize{}1\,kg} & {\scriptsize{}1\,kg} & {\scriptsize{}1\,kg}\tabularnewline
{\scriptsize{}Scale} & {\scriptsize{}-} & {\scriptsize{}1} & {\scriptsize{}1} & {\scriptsize{}1} & {\scriptsize{}1}\tabularnewline
{\scriptsize{}CAPEX-scaled $\left(\unitfrac{\text{€}}{kg}\right)$} & {\scriptsize{}-} & {\scriptsize{}450} & {\scriptsize{}3.5} & {\scriptsize{}13.31} & {\scriptsize{}10}\tabularnewline
{\scriptsize{}OPEX (\%) \citep{Reuss2017a}} & {\scriptsize{}-} & {\scriptsize{}2} & {\scriptsize{}2.5 \citep{Stolzenburg2014a}} & {\scriptsize{}2} & {\scriptsize{}2}\tabularnewline
{\scriptsize{}Depreciation period (a) \citep{Parks2014a}} & {\scriptsize{}-} & {\scriptsize{}20} & {\scriptsize{}30 \citep{Stolzenburg2014a}} & {\scriptsize{}20} & {\scriptsize{}20}\tabularnewline
{\scriptsize{}Pressure range (bar)} & {\scriptsize{}-} & {\scriptsize{}15 - 250} & {\scriptsize{}60 - 180} & {\scriptsize{}-} & {\scriptsize{}-}\tabularnewline
{\scriptsize{}Min. filling level (\%)\textsuperscript{a}} & {\scriptsize{}-} & {\scriptsize{}6} & {\scriptsize{}33.3} & {\scriptsize{}5} & {\scriptsize{}-}\tabularnewline
{\scriptsize{}Boil-off $\left(\unitfrac{\%}{d}\right)$ \citep{Bouwkamp2017a}} & {\scriptsize{}-} & {\scriptsize{}-} & {\scriptsize{}-} & {\scriptsize{}0.2} & {\scriptsize{}-}\tabularnewline
{\scriptsize{}Storage bypass possibility} & {\scriptsize{}-} & {\scriptsize{}yes} & {\scriptsize{}yes} & {\scriptsize{}-} & {\scriptsize{}-}\tabularnewline
\hline 
\end{tabular}
\par\end{centering}

\begin{spacing}{0.40000000000000002}
\centering{}%
\begin{tabular}{p{12.7cm}}
\textit{\tiny{}Abbreviations:}{\tiny{} cav.: cavern; (S): small-scale on-site supply chain; (L): large-scale supply chain}{\tiny \par}
{\tiny{}a: \vphantom{\unitfrac{kg}{h}}Calculated according to Boyle's law in order to maintain
the minimum pressure required. For the cavern, minimum pressure is
calculated dependent on the required amount of cushion gas.}\tabularnewline
\end{tabular}\end{spacing}
\end{table}

\par\end{center}


\begin{center}
\begin{table}[H]
\protect\caption{\label{tab: transporation auxiliaries (before)}Assumptions for different transportation preparation processes}

\begin{centering}
\begin{tabular}{llllll}
\hline 
 & {\scriptsize{}GH\textsubscript{{\scriptsize{}2}} (S) } & {\scriptsize{}GH\textsubscript{{\scriptsize{}2}} (L) \citep{Elgowainy2015a}} & {\scriptsize{}GH\textsubscript{{\scriptsize{}2}}\textsuperscript{{\scriptsize{}cav.}}
(L) \citep{Elgowainy2015a}} & {\scriptsize{}LH\textsubscript{{\scriptsize{}2}} (L)} & {\scriptsize{}LOHC (L)}\tabularnewline
\hline 
{\scriptsize{}Activity} & {\scriptsize{}-} & {\scriptsize{}compression} & {\scriptsize{}compression} & \multicolumn{2}{c}{{\scriptsize{}overflow/pumping}}\tabularnewline
{\scriptsize{}CAPEX-base (€)} & {\scriptsize{}-} & {\scriptsize{}6000} & {\scriptsize{}6000} & {\scriptsize{}-} & {\scriptsize{}-}\tabularnewline
{\scriptsize{}CAPEX-comparison} & {\scriptsize{}-} & {\scriptsize{}$\unit[1]{kW_{el}}$} & {\scriptsize{}$\unit[1]{kW_{el}}$} & {\scriptsize{}-} & {\scriptsize{}-}\tabularnewline
{\scriptsize{}Scale} & {\scriptsize{}-} & {\scriptsize{}1} & {\scriptsize{}1} & {\scriptsize{}-} & {\scriptsize{}-}\tabularnewline
{\scriptsize{}Ref.-Capacity $\left(\unitfrac{kg}{h}\right)$} &  & {\scriptsize{}720} & {\scriptsize{}720} & {\scriptsize{}-} & {\scriptsize{}-}\tabularnewline
{\scriptsize{}CAPEX-scaled $\left(\unitfrac{\text{€}}{kg}\right)$\textsuperscript{a}} & {\scriptsize{}-} & {\scriptsize{}13,784} & {\scriptsize{}6,530} & {\scriptsize{}-} & {\scriptsize{}-}\tabularnewline
{\scriptsize{}OPEX (\%)} & {\scriptsize{}-} & {\scriptsize{}4} & {\scriptsize{}4} & {\scriptsize{}-} & {\scriptsize{}-}\tabularnewline
{\scriptsize{}Depreciation period (a)} & {\scriptsize{}-} & {\scriptsize{}15} & {\scriptsize{}15} & {\scriptsize{}-} & {\scriptsize{}-}\tabularnewline
{\scriptsize{}Min. Pressure in (bar)} & {\scriptsize{}-} & {\scriptsize{}15} & {\scriptsize{}60} & {\scriptsize{}-} & {\scriptsize{}-}\tabularnewline
{\scriptsize{}Pressure out (bar)} & {\scriptsize{}-} & {\scriptsize{}250} & {\scriptsize{}250} & {\scriptsize{}-} & {\scriptsize{}-}\tabularnewline
{\scriptsize{}Compression stages} & {\scriptsize{}-} & {\scriptsize{}2} & {\scriptsize{}2} & {\scriptsize{}-} & {\scriptsize{}-}\tabularnewline
{\scriptsize{}Elec. demand $\left(\unitfrac{kWh}{kg}\right)$} & {\scriptsize{}-} & {\scriptsize{}2.297} & {\scriptsize{}1.088} & {\scriptsize{}-} & {\scriptsize{}-}\tabularnewline
{\scriptsize{}Losses (\%)} & {\scriptsize{}-} & {\scriptsize{}0.5} & {\scriptsize{}0.5} & {\scriptsize{}-} & {\scriptsize{}-}\tabularnewline
\hline 
\end{tabular}
\par\end{centering}

\begin{spacing}{0.40000000000000002}
\centering{}%
\begin{tabular}{p{13.3cm}}
\textit{\tiny{}Abbreviations:}{\tiny{} cav.: cavern; (S): small-scale on-site supply chain; (L): large-scale supply chain}{\tiny \par}
{\tiny{}a: \unit[720]{kg/h} is equal to the trailer capacity.
Thus, every compressor is required to have the capacity to load one
trailer per hour.}\tabularnewline
\end{tabular}\end{spacing}
\end{table}

\par\end{center}


\begin{center}
\begin{table}[H]
\protect\caption{\label{tab: transporation processes}Assumptions for different transportation processes}

\begin{centering}
\begin{tabular}{lllll}
\hline 
 & {\scriptsize{}All \citep{Teichmann2012a}} & {\scriptsize{}GH\textsubscript{{\scriptsize{}2}} (L) \citep{US-DOE-2015a}} & {\scriptsize{}LH\textsubscript{{\scriptsize{}2}} (L) \citep{US-DOE-2015a}} & {\scriptsize{}LOHC (L) \citep{Reuss2017a}}\tabularnewline
\hline 
{\scriptsize{}Function} & {\scriptsize{}tractor} & {\scriptsize{}trailer} & {\scriptsize{}trailer} & {\scriptsize{}trailer}\tabularnewline
{\scriptsize{}CAPEX (€)\textsuperscript{a,\,b}} & {\scriptsize{}223,031} & {\scriptsize{}518,400} & {\scriptsize{}865,260} & {\scriptsize{}150,000}\tabularnewline
{\scriptsize{}Capacity (kg)} & {\scriptsize{}-} & {\scriptsize{}720} & {\scriptsize{}4,554} & {\scriptsize{}1,800}\tabularnewline
{\scriptsize{}Net capacity (kg)\textsuperscript{c}} & {\scriptsize{}-} & {\scriptsize{}676.8} & {\scriptsize{}4,326} & {\scriptsize{}1,620}\tabularnewline
{\scriptsize{}CAPEX-net $\left(\unitfrac{\text{€}}{kg}\right)$} & {\scriptsize{}-} & {\scriptsize{}763.93} & {\scriptsize{}190} & {\scriptsize{}92.59}\tabularnewline
{\scriptsize{}OPEX (\%)} & {\scriptsize{}12} & {\scriptsize{}2} & {\scriptsize{}2} & {\scriptsize{}2}\tabularnewline
{\scriptsize{}Depreciation period (a) \citep{Teichmann2012a}} & {\scriptsize{}12} & {\scriptsize{}12} & {\scriptsize{}12} & {\scriptsize{}12}\tabularnewline
{\scriptsize{}Losses $\left(\unitfrac{\%}{d}\right)$ \citep{Bouwkamp2017a}} & {\scriptsize{}-} & {\scriptsize{}-} & {\scriptsize{}0.6} & {\scriptsize{}-}\tabularnewline
{\scriptsize{}(Un-)/Loading time (h)} & {\scriptsize{}-} & {\scriptsize{}1 / 1} & {\scriptsize{}1 / 1} & {\scriptsize{}1 / 1}\tabularnewline
\hline 
\end{tabular}
\par\end{centering}

\begin{spacing}{0.40000000000000002}
\centering{}%
\begin{tabular}{p{12cm}}
\textit{\tiny{}Abbreviations:}{\tiny{} (L): large-scale supply chain}{\tiny \par}
{\tiny{}a: \vphantom{\unitfrac{kg}{h}}CAPEX adjusted for a lifetime of~$12$~years with an interest
rate of~$4$\,\%.}{\tiny \par}

{\tiny{}b: The average fuel consumption of a tractor is assumed to
be~\unit[35]{L/100\,km}~\citep{Teichmann2012a}. Moreover,
we assume a price of~\unit[1.30]{€/L} for diesel and
an hourly wage of drivers of~\unit[35]{€}. Fuel is not covered by the CO$_2$
tax.}{\tiny \par}

{\tiny{}c: For GH\textsubscript{{\tiny{}2}}, net-capacity is determined
by the required outlet pressure. $5$\,\% of LH\textsubscript{{\tiny{}2}}
remain in the trailer to avoid heating up of the trailer-tank. For
LOHC, a maximum discharge-depth of~$90$\,\% is assumed~\citep{Eypasch2017a}. Thus, transportation capacity of actually usable hydrogen is below the total amount of bound hydrogen. For all other processes, issues linked to a discharge-depth below~$100$\,\% are ignored either because the effect on costs is negligible (storage, degradation) or because we assume a heat-recovery system being installed (dehydrogenation).}\tabularnewline
\end{tabular}\end{spacing}
\end{table}

\par\end{center}


\begin{center}
\begin{table}[H]
\protect\caption{\label{tab:first fillling site auxiliaries}Assumptions for different filling storage preparation processes (1\protect\textsuperscript{st} stage)}

\begin{centering}
\begin{tabular}{lllll}
\hline 
 & {\scriptsize{}GH\textsubscript{{\scriptsize{}2}} (S)} & {\scriptsize{}GH\textsubscript{{\scriptsize{}2}}(L) \citep{Elgowainy2015a}} & {\scriptsize{}LH\textsubscript{{\scriptsize{}2}} (L)} & {\scriptsize{}LOHC (L)}\tabularnewline
\hline 
{\scriptsize{}Activity} & {\scriptsize{}-} & {\scriptsize{}compression} & \multicolumn{2}{c}{{\scriptsize{}overflow/pumping}}\tabularnewline
{\scriptsize{}CAPEX-base (€)} & {\scriptsize{}-} & {\scriptsize{}40,035} & {\scriptsize{}-} & {\scriptsize{}-}\tabularnewline
{\scriptsize{}CAPEX-comparison} & {\scriptsize{}-} & {\scriptsize{}$\unit[1]{kW_{el}}$} & {\scriptsize{}-} & {\scriptsize{}-}\tabularnewline
{\scriptsize{}Scale} & {\scriptsize{}-} & {\scriptsize{}0.6038} & {\scriptsize{}-} & {\scriptsize{}-}\tabularnewline
{\scriptsize{}Ref.-Capacity $\left(\unitfrac{kg}{h}\right)$} &  & {\scriptsize{}676.8} & {\scriptsize{}-} & {\scriptsize{}-}\tabularnewline
{\scriptsize{}CAPEX-scaled $\left(\unitfrac{\text{€}}{kg}\right)$} & {\scriptsize{}-} & {\scriptsize{}4,744} & {\scriptsize{}-} & {\scriptsize{}-}\tabularnewline
{\scriptsize{}OPEX (\%)} & {\scriptsize{}-} & {\scriptsize{}4} & {\scriptsize{}-} & {\scriptsize{}-}\tabularnewline
{\scriptsize{}Depreciation period (a)} & {\scriptsize{}-} & {\scriptsize{}15} & {\scriptsize{}-} & {\scriptsize{}-}\tabularnewline
{\scriptsize{}Pressure in (bar)} & {\scriptsize{}-} & {\scriptsize{}15} & {\scriptsize{}-} & {\scriptsize{}-}\tabularnewline
{\scriptsize{}Pressure out (bar)} & {\scriptsize{}-} & {\scriptsize{}250} & {\scriptsize{}-} & {\scriptsize{}-}\tabularnewline
{\scriptsize{}Compression stages\citep{Reuss2017a}} & {\scriptsize{}-} & {\scriptsize{}4} & {\scriptsize{}-} & {\scriptsize{}-}\tabularnewline
{\scriptsize{}Elec. demand $\left(\unitfrac{kWh}{kg}\right)$} & {\scriptsize{}-} & {\scriptsize{}2.105} & {\scriptsize{}-} & {\scriptsize{}-}\tabularnewline
{\scriptsize{}Constraint $\left(\unitfrac{trailers}{h}\right)$\textsuperscript{a}} & {\scriptsize{}-} & {\scriptsize{}1} & {\scriptsize{}1} & {\scriptsize{}1}\tabularnewline
{\scriptsize{}Losses (\%)} & {\scriptsize{}-} & {\scriptsize{}0.5} & {\scriptsize{}2.5} & {\scriptsize{}-}\tabularnewline
\hline 
\end{tabular}
\par\end{centering}

\begin{spacing}{0.40000000000000002}
\centering{}%
\begin{tabular}{p{10.4cm}}
\textit{\tiny{}Abbreviations:}{\tiny{} (S): small-scale on-site
supply chain; (L): large-scale supply chain}{\tiny \par}
{\tiny{}a: \vphantom{\unitfrac{kg}{h}}Own assumption to avoid congestion at the filling station.}\tabularnewline
\end{tabular}\end{spacing}
\end{table}

\par\end{center}


\begin{center}
\begin{table}[H]
\protect\caption{\label{tab: first filling site storage}Assumptions for different storage technologies (1\protect\textsuperscript{st} stage)}

\begin{centering}
\begin{tabular}{lllll}
\hline 
 & {\scriptsize{}GH\textsubscript{{\scriptsize{}2}} (S)} & {\scriptsize{}GH\textsubscript{{\scriptsize{}2}} (L) \citep{Parks2014a}} & {\scriptsize{}LH\textsubscript{{\scriptsize{}2}} (C) \citep{US-DOE-2015a}} & {\scriptsize{}LOHC (L) \citep{Reuss2017a}}\tabularnewline
\hline 
{\scriptsize{}CAPEX-base (€)} & {\scriptsize{}-} & {\scriptsize{}450} & {\scriptsize{}13.31} & {\scriptsize{}10}\tabularnewline
{\scriptsize{}CAPEX-comparison} & {\scriptsize{}-} & {\scriptsize{}1\,kg} & {\scriptsize{}1\,kg} & {\scriptsize{}1\,kg}\tabularnewline
{\scriptsize{}Scale} & {\scriptsize{}-} & {\scriptsize{}1} & {\scriptsize{}1} & {\scriptsize{}1}\tabularnewline
{\scriptsize{}CAPEX-scaled $\left(\unitfrac{\text{€}}{kg}\right)$} & {\scriptsize{}-} & {\scriptsize{}450} & {\scriptsize{}13.31} & {\scriptsize{}10}\tabularnewline
{\scriptsize{}OPEX (\%) \citep{Reuss2017a}} & {\scriptsize{}-} & {\scriptsize{}2} & {\scriptsize{}2} & {\scriptsize{}2}\tabularnewline
{\scriptsize{}Depreciation period (a) \citep{Parks2014a}} & {\scriptsize{}-} & {\scriptsize{}20} & {\scriptsize{}20} & {\scriptsize{}20}\tabularnewline
{\scriptsize{}Pressure range (bar)} & {\scriptsize{}-} & {\scriptsize{}15 - 250} & {\scriptsize{}-} & {\scriptsize{}-}\tabularnewline
{\scriptsize{}Min. filling level (\%)\textsuperscript{a}} & {\scriptsize{}-} & {\scriptsize{}6} & {\scriptsize{}5} & {\scriptsize{}-}\tabularnewline
{\scriptsize{}Boil-off $\left(\unitfrac{\%}{d}\right)$ \citep{Bouwkamp2017a}} & {\scriptsize{}-} & {\scriptsize{}-} & {\scriptsize{}0.4} & {\scriptsize{}-}\tabularnewline
{\scriptsize{}Storage bypass possibility} & {\scriptsize{}-} & {\scriptsize{}yes} & {\scriptsize{}-} & {\scriptsize{}-}\tabularnewline
\hline 
\end{tabular}
\par\end{centering}

\begin{spacing}{0.40000000000000002}
\centering{}%
\begin{tabular}{p{12.2cm}}
\textit{\tiny{}Abbreviations:}{\tiny{}  (S): small-scale on-site
supply chain; (L): large-scale supply chain}{\tiny \par}
{\tiny{}a: \vphantom{\unitfrac{kg}{h}}Calculated according to Boyle's law in order to maintain
the minimum pressure required.}\tabularnewline
\end{tabular}\end{spacing}
\end{table}

\par\end{center}


\begin{center}
\begin{table}[H]
\protect\caption{\label{tab: second fillling site auxiliaries-1}Assumptions for different filling storage preparation processes (2\protect\textsuperscript{nd} stage)}

\centering{}%
\begin{turn}{+90}
\begin{tabular}{lllllll}
\hline 
 & {\scriptsize{}GH\textsubscript{{\scriptsize{}2}} (S)} & {\scriptsize{}GH\textsubscript{{\scriptsize{}2}} (L)} & {\scriptsize{}LH\textsubscript{{\scriptsize{}2}} (L)} & {\scriptsize{}LH\textsubscript{{\scriptsize{}2}} (L)} & {\scriptsize{}LOHC (L)} & {\scriptsize{}LOHC (L)}\tabularnewline
 & {\scriptsize{}\citep{Elgowainy2015a}} & {\scriptsize{}\citep{Elgowainy2015a}} & {\scriptsize{}\citep{Nexant2008a,Elgowainy2015a}} & {\scriptsize{}\citep{Nexant2008a,Elgowainy2015a}} & {\scriptsize{}\citep{McClaine2015a,Teichmann2012a,Reuss2017a,Eypasch2017a,Muller2015a}} & {\scriptsize{}\citep{Elgowainy2015a}}\tabularnewline[\doublerulesep]
\hline 
{\scriptsize{}Activity} & {\scriptsize{}compression} & {\scriptsize{}compression} & {\scriptsize{}cryo-compr.-pump} & {\scriptsize{}evaporation} & {\scriptsize{}dehydrogenation} & {\scriptsize{}compression}\tabularnewline
{\scriptsize{}CAPEX-base (€)} & {\scriptsize{}40,035} & {\scriptsize{}40,035} & {\scriptsize{}567.1\,$\unitfrac{\text{€}}{kg}$} & {\scriptsize{}900.9\,$\unitfrac{\text{€}}{kg}$} & {\scriptsize{}55,707} & {\scriptsize{}40,035}\tabularnewline
 &  &  & {\scriptsize{}+ 11,565\,€} & {\scriptsize{}+ 2,389\,€} &  & \tabularnewline
{\scriptsize{}CAPEX-comparison} & {\scriptsize{}$\unit[1]{kW_{el}}$} & {\scriptsize{}$\unit[1]{kW_{el}}$} & {\scriptsize{}1\,kg} & {\scriptsize{}1\,kg} & {\scriptsize{}1\,kg} & {\scriptsize{}$\unit[1]{kW_{el}}$}\tabularnewline
{\scriptsize{}Scale} & {\scriptsize{}0.6038} & {\scriptsize{}0.6038} & {\scriptsize{}1} & {\scriptsize{}1} & {\scriptsize{}2/3} & {\scriptsize{}0.6038}\tabularnewline
{\scriptsize{}Ref.-Capacity $\left(\unitfrac{kg}{h}\right)$} & {\scriptsize{}45} & {\scriptsize{}45} & {\scriptsize{}45} & {\scriptsize{}45} & {\scriptsize{}45} & {\scriptsize{}45}\tabularnewline
{\scriptsize{}CAPEX-scaled $\left(\unitfrac{\text{€}}{kg}\right)$} & {\scriptsize{}17,014} & {\scriptsize{}19,070} & {\scriptsize{}824.1} & {\scriptsize{}954} & {\scriptsize{}15,662} & {\scriptsize{}22,220}\tabularnewline
{\scriptsize{}OPEX (\%)} & {\scriptsize{}4} & {\scriptsize{}4} & {\scriptsize{}4} & {\scriptsize{}1} & {\scriptsize{}4} & {\scriptsize{}4}\tabularnewline
{\scriptsize{}Depreciation period (a)} & {\scriptsize{}10} & {\scriptsize{}10} & {\scriptsize{}10} & {\scriptsize{}10} & {\scriptsize{}20} & {\scriptsize{}10}\tabularnewline
{\scriptsize{}Pressure in (bar)} & {\scriptsize{}30} & {\scriptsize{}15} & {\scriptsize{}2} & {\scriptsize{}-} & {\scriptsize{}-} & {\scriptsize{}5 \citep{Eypasch2017a}}\tabularnewline
{\scriptsize{}Pressure out (bar)} & {\scriptsize{}950} & {\scriptsize{}950} & {\scriptsize{}-} & {\scriptsize{}950} & {\scriptsize{}5} & {\scriptsize{}950}\tabularnewline
{\scriptsize{}Compression stages \citep{Reuss2017a}} & {\scriptsize{}4} & {\scriptsize{}4} & {\scriptsize{}-} & {\scriptsize{}-} & {\scriptsize{}-} & {\scriptsize{}4}\tabularnewline
{\scriptsize{}Elec. demand $\left(\unitfrac{kWh}{kg}\right)$} & {\scriptsize{}2.947} & {\scriptsize{}3.559} & {\scriptsize{}0.1 \citep{Reuss2017a}} & {\scriptsize{}0.6 \citep{Reuss2017a}} & {\scriptsize{}-} & {\scriptsize{}4.585}\tabularnewline
{\scriptsize{}Heat demand $\left(\unitfrac{kWh}{kg}\right)$\textsuperscript{a}} & {\scriptsize{}-} & {\scriptsize{}-} & {\scriptsize{}-} & {\scriptsize{}-} & {\scriptsize{}9.1} & {\scriptsize{}-}\tabularnewline
{\scriptsize{}Losses (\%)} & {\scriptsize{}0.5} & {\scriptsize{}0.5} & {\scriptsize{}-} & {\scriptsize{}-} & {\scriptsize{}1} & {\scriptsize{}0.5}\tabularnewline
\hline 
\noalign{\vskip-0.1cm}
\multicolumn{7}{l}{{{\hskip-0.1cm}\textit{\tiny{}Abbreviations:}{\tiny{}  (S): small-scale on-site supply chain; (L): large-scale supply chain}{\tiny \par}}}\tabularnewline
\noalign{\vskip-0.25cm}
\multicolumn{7}{l}{{\tiny{}{\hskip-0.1cm}a: \unit[8.9]{kWh/kg} \citep{Reuss2017a,Muller2015a}
corrected for~$97.5$\,\% heat exchanger efficiency as described in~\citep{Eypasch2017a}.}}\tabularnewline
\end{tabular}
\end{turn}
\end{table}

\par\end{center}


\begin{center}
\begin{table}[H]
\protect\caption{\label{tab: second filling site storage}Assumptions for different storage technologies (2\protect\textsuperscript{nd} stage)}

\begin{centering}
\begin{tabular}{ll}
\hline 
 & {\scriptsize{}All \citep{US-DOE-2015a}}\tabularnewline
\hline 
{\scriptsize{}CAPEX-base (€)} & {\scriptsize{}600}\tabularnewline
{\scriptsize{}CAPEX-comparison} & {\scriptsize{}1\,kg}\tabularnewline
{\scriptsize{}Scale} & {\scriptsize{}1}\tabularnewline
{\scriptsize{}CAPEX-scaled $\left(\unitfrac{\text{€}}{kg}\right)$} & {\scriptsize{}600}\tabularnewline
{\scriptsize{}OPEX (\%)} & {\scriptsize{}2 \citep{Reuss2017a}}\tabularnewline
{\scriptsize{}Depreciation period (a)} & {\scriptsize{}20 \citep{Parks2014a}}\tabularnewline
{\scriptsize{}Pressure range (bar)} & {\scriptsize{}700 - 950}\tabularnewline
{\scriptsize{}Min. filling level (\%)\textsuperscript{a}} & {\scriptsize{}74}\tabularnewline
\hline 
\end{tabular}
\par\end{centering}

\begin{spacing}{0.40000000000000002}
\centering{}%
\begin{tabular}{p{5cm}}
{\tiny{}a: Calculated according to Boyle's law in order to maintain
the minimum pressure required.}\tabularnewline
\end{tabular}\end{spacing}
\end{table}

\par\end{center}


\begin{center}
\begin{table}[H]
\protect\caption{\label{tab: filling station equipment}Assumptions for filling station equipment}

\begin{centering}
\begin{tabular}{lll}
\hline 
 & {\scriptsize{}Refrigeration \citep{Elgowainy2015a}} & {\scriptsize{}Dispenser \citep{Elgowainy2015a}}\tabularnewline
\hline 
{\scriptsize{}CAPEX-base $\left(\nicefrac{\text{€}}{\textrm{pc.}}\right)$
\citep{US-DOE-2015a}} & {\scriptsize{}70,000} & {\scriptsize{}60,000}\tabularnewline
{\scriptsize{}OPEX (\%)} & {\scriptsize{}2} & {\scriptsize{}1}\tabularnewline
{\scriptsize{}Depreciation period (a)} & {\scriptsize{}15} & {\scriptsize{}10}\tabularnewline
{\scriptsize{}Elec. demand $\left(\unitfrac{kWh}{kg}\right)$} & {\scriptsize{}0.325} & {\scriptsize{}-}\tabularnewline
{\scriptsize{}Max. temperature ($\unit{\text{\textdegree C}}$)\textsuperscript{a}} & {\scriptsize{}-40} & {\scriptsize{}-40}\tabularnewline
\hline 
\end{tabular}
\par\end{centering}

\begin{spacing}{0.40000000000000002}
\centering{}%
\begin{tabular}{p{8.4cm}}
{\tiny{}a: Hydrogen is dispensed to cars in gaseous form at~$\unit[700]{bar}$ and pre-cooled to $\unit[-40]{\text{\textdegree C}}$ in order to guarantee short filling times~\citep{Elgowainy2015a}.}\tabularnewline
\end{tabular}\end{spacing}
\end{table}

\par\end{center}


\begin{center}
\begin{table}[H]
\protect\caption{\label{tab: sensitivity: mass storage for dec}Sensitivity: mass storage for small-scale on-site electrolysis}

\centering{}%
\begin{tabular}{lll}
\hline 
 & {\scriptsize{}GH\textsubscript{{\scriptsize{}2}} (S) \citep{Elgowainy2015a}} & {\scriptsize{}GH\textsubscript{{\scriptsize{}2}} (S) \citep{Elgowainy2015a}}\tabularnewline
\hline 
{\scriptsize{}Activity} & {\scriptsize{}compression (mass storage)} & {\scriptsize{}compression (high-pressure storage)}\tabularnewline
{\scriptsize{}CAPEX-base (€)} & {\scriptsize{}40,035} & {\scriptsize{}40,035}\tabularnewline
{\scriptsize{}CAPEX-comparison} & {\scriptsize{}$\unit[1]{kW_{el}}$} & {\scriptsize{}$\unit[1]{kW_{el}}$}\tabularnewline
{\scriptsize{}Scale} & {\scriptsize{}0.6038} & {\scriptsize{}0.6038}\tabularnewline
{\scriptsize{}Ref.-Capacity $\left(\unitfrac{kg}{h}\right)$} & {\scriptsize{}45} & {\scriptsize{}45}\tabularnewline
{\scriptsize{}CAPEX-scaled $\left(\unitfrac{\text{€}}{kg}\right)$} & {\scriptsize{}11,972} & {\scriptsize{}17,014}\tabularnewline
{\scriptsize{}OPEX (\%)} & {\scriptsize{}4} & {\scriptsize{}4}\tabularnewline
{\scriptsize{}Depreciation period (a)} & {\scriptsize{}15} & {\scriptsize{}10}\tabularnewline
{\scriptsize{}Pressure in (bar)} & {\scriptsize{}30} & {\scriptsize{}30}\tabularnewline
{\scriptsize{}Pressure out (bar)} & {\scriptsize{}250} & {\scriptsize{}950}\tabularnewline
{\scriptsize{}Compression stages\citep{Reuss2017a}} & {\scriptsize{}4} & {\scriptsize{}4}\tabularnewline
{\scriptsize{}Elec. demand $\left(\unitfrac{kWh}{kg}\right)$} & {\scriptsize{}1.654} & {\scriptsize{}2.947}\tabularnewline
{\scriptsize{}Losses (\%)} & {\scriptsize{}0.5} & {\scriptsize{}0.5}\tabularnewline
\hline 
\end{tabular}

\begin{spacing}{0.40000000000000002}
\centering{}%
\begin{tabular}{p{12.5cm}}
\textit{\tiny{}Abbreviations:}{\tiny{} (S): small-scale on-site supply chain}{\tiny \par}
\tabularnewline
\end{tabular}\end{spacing}

\end{table}

\par\end{center}


\renewcommand{\refname}{SI References}
\addcontentsline{toc}{section}{\refname}
\putbib[references-repository-p2x]
\end{bibunit}

\end{document}